\documentclass[12pt,manuscript]{aastex}

\usepackage{graphics,graphicx}

\begin{document} 


\title{Unearthing Foundations of a Cosmic Cathedral: Searching the
  Stars for M33's Halo}

\author{Robert Cockcroft}
\affil{Department of Physics and Astronomy, McMaster University, Hamilton, Ontario, L8S 4M1, Canada}
\email{cockcroft@physics.mcmaster.ca}

\author{Alan W. McConnachie}
\affil{NRC Herzberg Institute of Astrophysics, 5071 West Saanich Road, Victoria, British Columbia, V9E 2E7, Canada}
\email{alan.mcconnachie@nrc-cnrc.gc.ca}

\author{William E. Harris}
\affil{Department of Physics and Astronomy, McMaster University, Hamilton, Ontario, L8S 4M1, Canada}
\email{harris@physics.mcmaster.ca}

\author{Rodrigo Ibata}
\affil{Observatoire Astronomique, Université de Strasbourg, CNRS, 11, rue de l'Université, F-67000 Strasbourg, France}
\email{ibata@astro.u-strasbg.fr}

\author{Mike J. Irwin}
\affil{Institute of Astronomy, University of Cambridge, Madingley Road, Cambridge CB3 0HA, UK}
\email{mike@ast.cam.ac.uk}

\author{Annette M. N. Ferguson}
\affil{Institute for Astronomy, University of Edinburgh, Blackford Hill, Edinburgh, EH9 3HJ, UK}
\email{ferguson@roe.ac.uk}

\author{Mark A. Fardal}
\affil{Department of Astronomy, University of Massachusetts, 710 North Pleasant Street, Amherst, MA, USA}
\email{fardal@astro.umass.edu}

\author{Arif Babul}
\affil{Department of Physics and Astronomy, University of Victoria, Victoria, BC, V8P 1A1, Canada}
\email{babul@uvic.ca}

\author{Scott C. Chapman}
\affil{Institute of Astronomy, University of Cambridge, Madingley Road, Cambridge CB3 0HA, UK}
\email{schapman@ast.cam.ac.uk}

\author{Geraint F. Lewis}
\affil{Institute of Astronomy, School of Physics, A28, University of Sydney, NSW 2006, Australia}
\email{gfl@physics.usyd.edu.au}

\author{Nicolas F. Martin}
\affil{Max-Planck-Institut für Astronomie, Königstuhl 17, D-69117 Heidelberg, Germany}
\email{martin@mpia.de}

\and

\author{Thomas H. Puzia}
\affil{Department of Astronomy and Astrophysics, Pontificia Universidad Católica de Chile, Av. Vicuna Mackenna 4860, 7820436 Macul, Santiago, Chile}
\email{tpuzia@gmail.com}


\begin{abstract}

We use data from the
Pan-Andromeda Archaeological Survey (PAndAS) to search for evidence of
an extended halo component belonging to M33 (the Triangulum Galaxy).
We identify a population
of red giant branch (RGB) stars at large radii from M33's disk whose
connection to the recently discovered extended ``disk 
substructure'' is ambiguous, and which may represent a ``bona-fide''
halo component. After first correcting for contamination from the
Milky Way foreground population and misidentified background galaxies,
we average the radial density of RGB candidate stars over circular
annuli centered on the galaxy and away from the disk substructure.  We
find evidence of a low-luminosity,   
centrally concentrated component that is everywhere in our data
fainter than $\mu_{V} \sim $ 33 mag arcsec$^{-2}$. The scale length of
this feature is not well constrained by our data, but it appears to be
of order $r_{exp} \sim $ 20 kpc; there is weak evidence to
suggest it is not azimuthally symmetric. Inspection of the overall CMD for this
region that specifically clips out the disk 
substructure reveals that this residual RGB 
population is consistent with an old population with a photometric
metallicity of around [Fe/H] $\sim$ -2 dex, but some residual
contamination from the disk substructure appears to remain.  We
discuss the likelihood that our findings represent a bona-fide halo in
M33, rather than extended emission from the disk substructure.  We
interpret our findings in terms of an upper limit to M33's halo
that is a few percent of its total luminosity, although its
actual luminosity is likely much less.

\end{abstract}


\section{Introduction and Background}
\label{intro}

$\Lambda$ Cold Dark Matter ($\Lambda$CDM) cosmology predicts that
larger galaxies are built through the hierarchical merging of smaller
galaxies.  Infalling components are disrupted partially or entirely,
and part of that material forms the stellar halo of the larger galaxy.
Stellar halos therefore contain the remnants of past interactions
between galaxies, and their properties can indicate the approximate time,
size and frequency of past mergers (e.g., \citealt{2007ApJ...666...20P}).  

We can directly observe only a few relatively nearby halos in any great
detail due to their faint nature, and only a small number of halos are
directly observed through their resolved stars outside of the Local
Group. Due to their faintness, 
it is problematic to determine whether or not the halos are smooth
and/or symmetric.  There is likely a continuum of scenarios that we
observe between newly-accreted objects (creating streams, shells,
etc.; e.g., \citealt{2010AJ....140..962M}) and smooth halos, and our interpretation will depend on the time
since the accretion and the spatial resolution and depth of the
observations. The long dynamical timescales for structures outside of
the disk implies that they are long-lived \citep{1996ApJ...465..278J}.    

Outside of the Local Group, halo detections are extremely challenging
as it becomes more difficult with increasing distance to
distinguish the halo from other stellar components (e.g.,
\citealt{2002AJ....124.1328D, 2008ASPC..396..187D}) and even more so
in the absence of kinematical data \citep{2009AJ....138.1469B,
  2012MNRAS.419.1489B}.   Scattered light and non-stellar pollution of
counts also interfere with halo detections (e.g.,
\citealt{2008MNRAS.388.1521D}).  (The surface brightness 
detection limits generally needed are $\ge$ 7 magnitudes fainter than
the sky where the ``darkest'' skies (20th percentile), at Mauna Kea,
are fainter than $\mu_{V} \gtrsim$ 21.3.)\footnote{http://www.gemini.edu/sciops/telescopes-and-sites/observing-condition-constraints} 

Searches for halos around distant galaxies began with deep
observations around single galaxies using surface brightness
photometry and, as recent studies continue to
do, focussed on late-type edge-on galaxies (e.g.,
\citealt{1994Natur.370..441S, 1998ApJ...504L..23S,
  2004MNRAS.352L...6Z, 2005Ap.....48..221T, 2007IAUS..241..321B,
  2007ApJ...667L..49D, 2007IAUS..241..523S, 2009MNRAS.396.1231R,
  2007MNRAS.381..873M, 2010ApJ...714L..12M, 2011ApJS..195...18R}).
Each stellar component 
is revealed more easily in the cross-section rather than the face-on
view.  An alternative technique stacks many re-scaled images of galaxies
together before looking for a halo signal \citep{2004MNRAS.347..556Z,
  2008MNRAS.388.1521D, 2010MNRAS.405.2697B, 2011arXiv1112.1696Z,
  2011arXiv1102.0793Z}, again highlighting the difficulty of detecting
halos because of their extreme faintness.

Halos have also been observed around other types of galaxies, not
just late-type edge-on galaxies: for example, the Virgo Cluster's
central elliptical galaxy, M87 \citep{1997ApJ...490..664W}, nearby
starburst galaxies \citep{2011ApJ...736...24B, 2011AampA...530A..23R,
  2012Natur.482..192R}, the Leo elliptical NGC 3379
\citep{2007ApJ...666..903H}, and the giant elliptical NGC 5128
(Centaurus A; see \citealt{1983ApJ...272L...5M, 2002AJ....124.3144P,
  2011AampA...526A.123R}).  

$\S$ \ref{local_halos} provides a literature review of stellar halos
in the Local Group.
Details of the PAndAS observations around M33 are given in $\S$ \ref{obs}.  We
are ultimately concerned with identifying the RGB stars in the M33
 halo (if it exists).  However, we must exclude the regions associated
with the extended optical substructure surrounding the disk identified
in \cite{2009Natur.461...66M, 2010ApJ...723.1038M}, and we must also
correctly account for and subtract off the contribution from the
foreground Milky Way disk and halo components, and the background
galaxies misidentified as stars.   $\S$ \ref{methods} 
describes these corrections and exclusions as part of the analysis.
We discuss our results in $\S$ \ref{conclusions}, before summarizing
in $\S$ \ref{summary}.

\section{Halos of the Local Group Galaxies}
\label{local_halos}

\subsection{The Milky Way Galaxy}
\label{mw}
The Local Group provides the closest opportunity to study a stellar
halo but even the Milky Way (MW) is problematic to observe because of the
restrictions and biases associated with viewing our Galaxy from
within - although it has obviously been studied in depth (e.g.,
see the annual review by \citealt{2008A&ARv..15..145H}, and references therein).  Current seemingly contradictory evidence means it is unclear whether
the MW stellar halo is oblate, prolate or triaxial
(\citealt{2006JPhCS..47..195N}, \citealt{2011MNRAS.416.2903D}), 
although models of the dark matter (DM) halo seem to favour triaxiality
(\citealt{2009ApJ...703L..67L}, \citealt{2010ApJ...714..229L}).
Numerous detections of substructure beyond the stellar bulge and disk are
another reason that this ambiguity remains - substructure such as the
Sagittarius dwarf 
galaxy \citep{1994Natur.370..194I} and associated tidal streams
\citep{2001ApJ...547L.133I, 2002MNRAS.332..921I, 2003ApJ...599.1082M}, the
Monoceros ring \citep{2003MNRAS.340L..21I, 2003ApJ...588..824Y,
  2003ApJ...594L.119C}, overdensities in Canis Major
\citep{2004MNRAS.348...12M, 2004MNRAS.355L..33M} and Virgo
\citep{2001ApJ...554L..33V, 2002ApJ...569..245N,  2006MNRAS.368.1811X,
  2008ApJ...673..864J}, clouds in the Triangulum-Andromeda region
\citep{2004ApJ...615..732R, 2007ApJ...668L.123M} and the Hercules-Aquila region
\citep{2007ApJ...657L..89B}, and finally the Orphan
\citep{2006ApJ...645L..37G, 2006ApJ...642L.137B, 2007ApJ...658..337B} and
Cetus Polar \citep{2009ApJ...700L..61N} streams.  

There is growing evidence to suggest that the MW halo has a dual halo,
with the different components a result of their different formation
processes (e.g., \citealt{2000AJ....119.2843C, 2007Natur.450.1020C, 
  2008ApJ...678..865M, 2010ApJ...714..663D, 2012ApJ...746...34B}),
such as satellite accretion and in-situ formation (e.g.,
\citealt{2008ApJ...680..295B, 2009ApJ...703.2177S,
  2012ApJ...749...77S, 2009ApJ...702.1058Z, 2010ApJ...725.2312O,
  2012MNRAS.420.2245M}).    

The total (dark plus luminous) mass of the Galaxy within 300 kpc is
estimated to be in the range 0.7 $\leq M_{MW} \leq$ 3.4 x $10^{12}$
M$_{\odot}$ \citep{2009MNRAS.397.1990B, 2010MNRAS.406..264W}.  The MW
stellar halo luminosity, including all substructure, is estimated to
be of order $L_{MW,halo,V} \sim 10^{9} L_{\odot}$
(\citealt{1990AJ.....99..572C}; \citealt{2005ApJ...635..931B}, and
references therein), compared to the MW host luminosity, $L_{MW, host,
  V}$ = 2.1$^{+1.0}_{-0.6}$ x 10$^{10} L_{\odot}$
\citep{1997ApJ...483..103S}.  

\subsection{The Andromeda Galaxy}
\label{m31}
Observations of the Andromeda and Triangulum Galaxies (M31
and M33, respectively) are free from the problems inherent with
viewing the MW from within, but are still close
enough to resolve individual stars \citep{1986ApJ...305..591M,
  1986AJ.....92..292C} - and many ground-based studies are now also
resolving individual stars beyond the Local Group (e.g.,
\citealt{2009AJ....138.1469B, 2012MNRAS.419.1489B,
  2011ApJ...736...24B, 2011ApJ...738..150T}). However, progress on the
study of M31's full extended stellar halo has only 
come relatively recently - and the degree to which it is similar to
the MW's halo is still uncertain \citep{2006ApJ...648..389K,
  2007ApJ...671.1591I, 2008ApJ...689..958K}.  M31's total (dark plus
luminous) mass is estimated to be in the range 9.0 x $10^{11}
M_{\odot} < M_{M31} \lesssim$ 2 x $10^{12} M_{\odot}$ \citep{2000MNRAS.316..929E, 
  2006ApJ...653..255C, 2010MNRAS.406..264W}.  The lower limit is
determined from the kinematics of RGB stars out to 60 kpc with 99$\%$
confidence \citep{2006ApJ...653..255C}.  An alternative dynamical
mass that uses kinematics of M31's giant stream rather than those of
the satellite galaxies,
globular clusters, planetary nebulae or RGB stars, gives $M_{M31, R<125 kpc} =
7.5^{+2.5}_{-1.3}$ x $10^{11} M_{\odot}$ \citep{2004MNRAS.351..117I}.  M31 has a
total host luminosity of $L_{M31, host, V} \sim 2.6$ x $10^{10} L_{\odot}$ 
\citep{1999A&ARv...9..273V}, which is approximately 25$\%$ brighter
than the MW. Many photometric substructures have been revealed around M31
\citep{2001Natur.412...49I, 2002AJ....124.1452F, 2005ApJ...628L.105I, 
	2005ApJ...634..287I, 2007ApJ...671.1591I,
  2009Natur.461...66M}. \cite{2005ApJ...628L.105I} fit minor-axis profiles of a
de Vaucouleurs law out to a projected radius of  $\sim$20 kpc, and
beyond this a power law (index $\sim$ -2.3) or exponential (scale
length $\sim14$ kpc). \cite{2007ApJ...671.1591I} also find a smooth 
underlying component to which they fit a Hernquist profile (scale
length $\sim55$ kpc), and a power law (index $\sim$ -1.91 $\pm$ 0.11), similar
to the MW halo. If symmetric, \citeauthor {2007ApJ...671.1591I} estimate
that the total luminosity of the smooth halo is $L_{M31,halo, V}
\sim10^{9} L_{\odot}$, again similar to the MW halo.  More recently,
\cite{2011ApJ...739...20C} combine ground- and space-based data from
several sources and decompose the resulting composite luminosity
profile to find a halo component described by a power law (index
$\sim$ -2.5 $\pm$ 0.2).  M31's halo has
further similarities to that of the MW's in terms of 
metallicity and velocity dispersion \citep{2006ApJ...653..255C}.  Using a Keck/DEIMOS sample of $\sim$800 stars,
 \citeauthor{2006ApJ...653..255C} find a non-rotating metal-poor ([Fe/H] 
$\sim$ -1.4) smooth halo between 10 - 70 kpc with no metallicity
gradient underlying the metal-rich ([Fe/H] $\sim$ -0.9) rotating
extended component.  Along with a comparable dark matter halo mass, the similar
metallicities and dispersions are suggestive that the early formation
periods of both were also similar. 

\subsection{The Magellanic Clouds}
\label{lmc}
The Large Magellanic Cloud (LMC) is the fourth most massive Local Group
member.  It has a total mass of $M_{LMC} \approx 10^{10} M_{\odot}$
\citep{2002AJ....124.2639V, 2009MNRAS.395..342B}. The total
host luminosity is $L_{LMC,host,V} = 3.0$ x $10^{9} L_{\odot}$
\citep{2009MNRAS.395..342B}. It may also have a stellar 
halo.  A study of RR Lyrae stars, generally 
associated with old and metal-poor stellar populations, finds that
these stars have a large velocity dispersion of 53 $\pm$ 10 km s$^{-1}$
\citep{2003Sci...301.1508M, 2004ApJ...601L.151A}.  A photometric and
spectroscopic survey has also revealed individual RGB stars enveloping
the LMC out to large distances and consistent with a de Vaucouleurs
profile, suggestive of a classical halo \citep{2009IAUS..256...51M}.
If the LMC does have a stellar halo it seems somewhat surprising given
that another tracer of old and metal-poor populations - globular
clusters - show no evidence for a halo as they lie within a disk
region around the LMC \citep{1983ApJ...272..488F,
  1992AJ....103..447S}.  This inconsistency could be explained if the
GCs are accreted in earlier, more gas-rich merger events compared to
those that populate the halo with individual stars
\citep{2007MNRAS.380.1669B}.  

The stellar outskirts of the Small Magellanic Cloud (SMC) have recently been
studied through the MAgellanic Periphery Survey (MAPS;
\citealt{2011ApJ...733L..10N}) to reveal a population of
nearly azimuthally-symmetric RGB stars out to a radius of 
$\sim$11 kpc.  The profile of these stars are well fitted with an
exponential profile (scale length $\sim$1 kpc) out to $\sim$8 kpc,
with a shallower profile beyond (scale length $\sim$7 kpc) - the
latter of which the authors suggest could be a stellar halo or a
population of extratidal stars.

\subsection{The Triangulum Galaxy}
\label{m33}
M33, the Triangulum Galaxy, is the third most massive galaxy in the Local Group, with a mass
close to one tenth that of M31 (\citealt{2000MNRAS.311..441C} measure
M33's rotation curve out to 16 kpc, and find an implied
dark halo mass of $M_{M33} \gtrsim 5$ x $10^{10} M_{\odot}$).  M33 has
a total host luminosity of $L_{M33, host, V} \sim 10^{9} L_{\odot}$ \citep{1991trcb.book.....D} and is more face-on
($i =$ 56$^{\circ}\pm1^{\circ}$, \citealt{1989AJ.....97...97Z}) than
M31 ($i =$ 77$^{\circ}$; e.g., \citealt{1973ApJ...181...61R,
  2006MNRAS.370.1499A}). It is classified as a SA(s)cd II-III galaxy
\citep{1991trcb.book.....D}, has little or no bulge component, a UV-
and X-ray bright nuclear cluster (e.g., \citealt{1981ApJ...246L..61L}, \citealt{1999ApJ...519L.135D} and \citealt{2004A&A...416..529F}), and
perhaps a bar \citep{2011MNRAS.414.3394J}, so most of the central
light is distributed over an exponential 
disk component (e.g., \citealt{1959ApJ...130..728D,
  1992AJ....103..104B, 1993ApJ...410L..79M, 1996ApJ...456..499M,
  2008ASPC..396...19C, 2010ApJ...723...54K, 2011MNRAS.414.3394J}).  The
distance estimates to M33 cover a wide range from 730 $\pm$ 168 kpc
\citep{2005Sci...307.1440B} to 964 $\pm$ 54 kpc
\citep{2006ApJ...652..313B}.  This disagreement appears to arise
because of the combination of the different techniques used, and also
perhaps due to inhomogeneous interstellar extinction in
M33. Consistent with \cite{2010ApJ...723.1038M}, we adopt a 
distance modulus for M33 of $(m-M_{0})=24.54\pm0.06$ (809$\pm$24 kpc;
\citealt{2004MNRAS.350..243M}) throughout this 
paper. This distance is based on the TRGB method, and is
consistent with our new findings (820$^{+20}_{-19}$ kpc;
\citealt{2012Conn}).  At 809 kpc, 1$^{\circ}$ corresponds to 14.1 kpc.

Is there an M33 halo akin to those found in the MW, M31 and possibly
the LMC? Previous claims of a detection for M33's halo have come from
various sources, in studies that resolve individual stars (RGB or RR
Lyrae stars), or globular clusters, which we briefly review here.  

\cite{1986ApJ...305..591M} used the Hale Telescope/PFUEI to observe
two fields both 7 kpc from the centre of each M31 and M33, along their
southeast minor axes.  By comparing the observed giant branches to
those for M92 ($<[M/H]> = -0.6$) and 47 Tuc ($<[M/H]> = -2.2$), they
inferred the presence of inner halos.  However,
\cite{2004AJ....128..224T} observed a field with WIYN/S2KB that
included the region studied by \citeauthor{1986ApJ...305..591M} and
found that the peak of the MDF showed a radial variation with a
gradient consistent with that of the inner disk region and {\it not}
of an inner halo. \cite{2002ApJ...564..712C} obtained WIYN/HYDRA
spectra for 107 of M33's star clusters.  These clusters, from a sample
with known integrated HST/WFPC2 colors, were selected to cover the entire
age range of M33's clusters (6 Myr to $>$13 Gyr;
\citealt{2001A&A...366..498C}).  \citeauthor{2002ApJ...564..712C}
observed a large velocity dispersion that, with Monte Carlo
simulations, suggested that old ($>1$ Gyr) clusters could be split
into two components which they associate with a 
disk population and the other with a halo.  Similarly,
\cite{2006AJ....132.1361S} observed 64 RR Lyrae variable stars (type
RRab) using HST/ACS to have a double peak in periods, again suggesting
two subpopulations: a disk and a halo component.  RR Lyrae stars are
only observed in populations older than 10 Gyr, and therefore stars in
these populations should be at least as old as the RR Lyraes.

Further hints of M33's halo also come from the following sources.
\cite{2007AJ....133.1125B, 2007AJ....133.1138B} inspected three
HST/ACS southeastern 
fields $\sim$20'-30' ($\sim$4.7-7.1 kpc, assuming a distance to M33 of
809 kpc; \citealt{2004MNRAS.350..243M}) from
M33's nucleus.  Mixed 
stellar populations were revealed in the CMDs, with an age range from
$<$ 100 Myr to a few Gyr. The authors compared synthetic populations
with the observed CMDs, and found that the mean age increased with
radius from $\sim6$ to $8$ Gyr, and the mean metallicity decreased from
$\sim$-0.7 to -0.9 dex. They concluded that while the fields are dominated
by a disk population, a halo component may also be present.
\cite{2008A&A...487..131C} used UKIRT/WFCAM near-infrared observations
of a 1.8 degree$^{2}$ region centred on M33 to look at the ratio of C-
to M-type AGB stars.  They also found metallicity and age gradients
such that the outer regions were more metal poor and a few Gyr older
than the central regions, in agreement with \cite{2007AJ....133.1125B,
  2007AJ....133.1138B}.  \cite{2007ASPC..374..239F} reviews results
for M31 and M33, and notes 
that while a halo-like component with a power-law structure was
proving elusive, the RGB narrows and becomes more metal poor beyond
$\sim$10 kpc. \cite{2008PASP..120..474T} used RGB and AGB 
star counts along M33's minor axis, and also observed a break in the
surface brightness profile at 11 kpc.  The profile at this region
appeared to change from an exponential to a power-law.
\cite{2011MNRAS.410..504B} use HST/ACS to observe two fields 9.1 kpc
and 11.6 kpc along M33's north major axis.  They find that the outer
field is old (7 $\pm$ 2 Gyr), moderately metal poor (mean $[M/H] \sim$ -0.8 $\pm$
0.3), and contained $\sim$30 times less stellar mass than the inner
field.  One of the interpretations that
\citeauthor{2011MNRAS.410..504B} discuss is that the outer field is a
transition zone from the outer disk to another structural component.
\cite{2011A&A...533A..91G} use Subaru/Suprime-Cam data 
with seven fields 10 $\lesssim$ $r$ $<$ 30 kpc from the centre of M33
in the NW and SE.  An exponential scale length of $\sim$7 kpc is found
for both regions, and these authors favour that this component is an
extended disk rather than a halo.

The previous section highlights that unambiguous detections of the various
galactic components - even for one of our closest neighbouring
galaxies - are still extremely difficult.  While studies of star
clusters and RR Lyrae stars show evidence for an old halo, studies of
individual stars in other subpopulations have only been marginally
conclusive.  Part of the problem was the limited coverage and/or
depth.

\cite{2006ApJ...647L..25M} undertook a spectroscopic
survey of RGB stars using Keck/DEIMOS.  Radial velocity distributions
of these stars were best-fit by three Gaussian components, which
\citeauthor{2006ApJ...647L..25M} interpreted as contributions from a
halo, a disk, and a component offset from the disk (which they suggested
could have been a stellar stream or another stellar
halo component). \cite{2007ApJ...671.1591I} extended the observations of
\cite{2007iuse.book..239F} with CFHT/MegaCam along the southeastern
corner of M31's halo out to M33's
centre. \citeauthor{2007ApJ...671.1591I} clearly saw the classical
disk of M33, and in addition revealed an extended component.
They fit a profile to 
the data between 1 and 4 degrees (the edge of the disk, and the point just
before where the profile starts rising again, respectively) and found the
exponential scale length to be 18 $\pm$ 1 kpc, or 55 $\pm$ 2 kpc using
a projected Hernquist model.  These scale lengths were
surprisingly as big as they found for M31, although
\citeauthor{2007ApJ...671.1591I} cautioned that without a full
panoramic view it was not possible to determine whether or not this
feature was a ``bona fide'' halo.  

Direct and unambiguous evidence for a stellar halo around M33
remains elusive but is the aim of this paper.  Any such component must
be quite faint.  We extend  
the work begun by \cite{2009Natur.461...66M, 2010ApJ...723.1038M} as
part of the Pan-Andromeda Archaeological Survey (PAndAS),
which itself built on previous surveys with the INT/WFC
\citep{2007iuse.book..239F} and CFHT/MegaCam
\citep{2007ApJ...671.1591I}.  Optical observations prior to PAndAS
suggested M33's disk had an undisturbed 
appearance - a view that persisted until relatively recently (e.g.,
\citealt{2007RMxAC..29...48S,
  2007iuse.book..239F}).  This implied that the disk had not been tidally
disrupted by either the MW or M31, and was seemingly discrepant when
compared to the radio detection of a warped gaseous disk component  
(e.g., \citealt{1976ApJ...204..703R, 2009ApJ...703.1486P}).  We note,
however, that the primary problem with the
\citeauthor{2007iuse.book..239F} data set was depth and that the
observations could only rule out the presence of substructure or a
halo with surface brightnesses brighter than $\mu_{V} \sim$31 mag
arcsec$^{-2}$. 

PAndAS observations covering the area around M33 with 
unprecedented combination of depth and coverage have revealed a vast
low surface brightness stellar substructure.  The S-shaped optical
warp of this substructure is generally aligned with the HI warp, and therefore
resolves the previous discrepancy. It now seems most likely that this
warp was the feature that was previously partially detected by
\cite{2006ApJ...647L..25M} (thereby casting doubt on the
previous interpretations from this kinematic study) and
\cite{2007ApJ...671.1591I}. Although the nature of the substructure is
still being investigated, the favoured interpretation for its origin
is that it is a disruption of the disk that was caused by a tidal
interaction with M31 as M33 orbits M31 \citep{2009Natur.461...66M,
  2010ApJ...723.1038M, Dubinski2012_in_prep}. Preliminary models can
reproduce the shape of the extended disk substructure, and also
satisfy M33's proper motion constraints \citep{2005Sci...307.1440B}.
Spectroscopic observations may provide further clues (see upcoming 
paper by \citealt{Trethewey2012_in_prep}). 

The depth of the PAndAS data allows us to test whether or not a
genuine halo component is observable in addition to the disk-like
warp. Given that the warped extended disk substructure is  
extremely faint and had eluded detection for so long, it would be
reasonable to expect any underlying component of a halo to also be
extremely faint.  Indeed, a faint extended stellar component is
hinted at beyond the extended disk substructure, and one possibility put
forward is that this is a halo component \citep{2010ApJ...723.1038M}. 

Since M33 is relatively low mass, would we expect it to have a
detectable halo? \cite{2007ApJ...666...20P} predict that a galaxy with
total mass $M \sim 10^{11} M_{\odot}$ will, {\it on average}, have a
halo that contributes $\le$1$\%$ of the total luminosity from the
galaxy.  Therefore, the expected total halo luminosity of M33 could be
as low as $L_{M33,halo} \le 10^{7} L_{\odot}$.  Their halo estimates
make no distinction between the smooth component and the substructure.
Here, we {\it define} the ``halo'' as the component or stellar
population in the outer regions of M33 that is not clearly associated
with the disk or the extended disk substructure identified by
\cite{2010ApJ...723.1038M}.  Note that we do not distinguish between
smooth or lumpy halos, similarly to \cite{2007ApJ...666...20P}. 


\section{Observations, Data Reduction and Calibration}
\label{obs}

We use data from the Pan-Andromeda Archaeological Survey (PAndAS;
\citealt{2009Natur.461...66M}) to observe 48 degree$^{2}$ around M33
with CFHT/MegaCam out to a projected radius of 50 kpc (cf. $r_{M33,virial}$ = 152 kpc, 
\citealt{2009ApJ...705..758M}).  The data has limiting
magnitudes for point-source detections of $g' \approx 25.5, i' \approx 24.5$ (AB magnitudes on the
SDSS scale) at S/N = 10 in subarcsecond seeing. MegaCam is composed of
36 individual CCDs, has a 0.96 x 0.94 degree field of view and a
resolution of 0.187'' pixel$^{-1}$.

\begin{figure}
  \begin{center}
    \includegraphics[width=90mm]{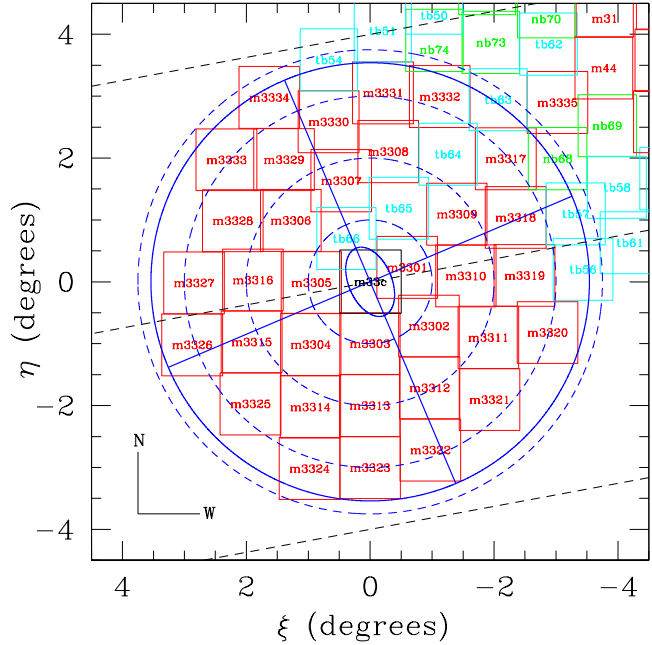}
  \end{center}
  \caption{A tangent-plane projection of the PAndAS fields around M33.
    The central field, m33c (black), is from the archive (the data is
    from 2003B and 2004B). All other fields with prefix m (red)
    were observed in 2008B. Fields with prefix nb (green) and tb
    (light blue) were observed in 2009B and 2010B, respectively. The
    dark blue solid ellipse marks the diameter (73 x 45 arcminutes) at
    which $\mu{_B} \approx$ 25 mag arcsec$^{-2}$
    \citep{1973UGC...C...0000N}.  The two perpendicular lines show the
    major and minor axes (the major axis is inclined 23 degrees from
    the vertical; 
    \citealt{1973UGC...C...0000N}).  The solid-line circle represents $r=$ 50
    kpc ($\approx 0.33$ $r_{M33,virial}$).  The concentric dashed-line circles mark radii of $r=$ 1, 2,
    3 and 3.75 degrees (14.1, 28.2, 42.4 and 53.0 kpc, respectively).  We
    assume a distance modulus of $(m-M_{0})=24.54\pm0.06$ (809$\pm$24
    kpc; \citealt{2004MNRAS.350..243M, 2005MNRAS.356..979M}). The
    three straight black dashed lines each represent one line of
    equivalent Galactic latitude ($b$ = -35.3, -31.3 and -27.3).
  }
  \label{frames}
\end{figure}

Figure \ref{frames} shows each location of the MegaCam
$\sim$square-degree images around M33 in a tangent-plane projection.
To be explicit, for our analysis we only consider the data for MegaCam
images within the annulus with radius 3.75 degrees centred on M33 as
shown in Figure \ref{frames}.  The MegaCam subexposures are dithered
in order to cover the small gaps, but not the large gaps\footnote{See
  http://www.cfht.hawaii.edu/Instruments/Imaging/Megacam/specsinformation.html}. In
the large gaps there may be fewer detections due to shallower 
depths. However, the area of the chip gaps is very small compared to
the overall MegaCam field and can be neglected for the purposes of
this analysis.  

The prefixes of the image labels in Figure \ref{frames} represent the
timeline of the observations: The central field, m33c, was observed
primarily in the observing semester 2004B and retrieved from the CFHT
archive, with some data from 2003B.  All other fields with 
prefix m were observed in 2008B. 
Fields with prefix nb were observed in 2009B.  Due to a failure of CCD
4 in the 2003B observing semester, the data from
\cite{2007ApJ...671.1591I} which extended the southeastern section of
M31's halo in a line to the centre of M33 was replaced with data from
2010B (prefix tb).  The ellipse in Figure \ref{frames} marks the $\mu{_B} \approx$ 25 mag arcsec$^{-2}$
\citep{1973UGC...C...0000N} contour of M33's disk, and the solid-line
cross represents the major and minor axes of M33, with the major axis
inclined 23 degrees to North \citep{1973UGC...C...0000N}.  The dashed
concentric circles represent radii at $r$ = 1, 2, 3 and 3.75 degrees
(14.1, 28.2, 42.4 and 53.0 kpc, respectively) from the centre of
M33.  The data within the annuli that they delineate will be used in
the analysis that follows.  M33 is approximately 31 degrees below the
central axis of the Milky Way disk, M33$_{(l, b)}$ = (133.61, -31.33)
degrees, compared with M31 which is about 21 degrees below, M31$_{(l,
b)}$ = (121.18, -21.57) degrees.  The three dashed lines in Figure
\ref{frames} are lines of equal Galactic latitude ($b$ = -35.3, -31.3
and -27.3 degrees).  

Pre-processing and reduction were undertaken with
Elixir\footnote{http://www.cfht.hawaii.edu/Instruments/Imaging/MegaPrime/dataprocessing.html}
by the CFHT team, and by the Cambridge Astronomical Survey Unit (CASU)
through a pipeline adapted for MegaCam images
\citep{2001NewAR..45..105I}, respectively. The reader is referred to
\cite{2009Natur.461...66M,   2010ApJ...723.1038M} and
\cite{2011ApJ...730..112C} for more details.   


\section{Analysis}
\label{methods}

Taking advantage of the wide coverage of the PAndAS data, we
deliberately seek direct evidence for M33's stellar halo in
this data, and expect it to be extremely faint, 
centrally-concentrated, and detectable via RGB
stars.  However, this low-luminosity component will be mixed with
stars from the M33 disk, M33 extended disk substructure surrounding the
disk, and the MW foreground (both its thick disk and halo), in
addition to background galaxies misidentified as stars.  Our technique involves
statistically removing the MW foreground stars and background
galaxies, excluding the regions identified as belonging to M33's
extended disk substructure, and seeing what signal remains.  

\begin{figure}
  \begin{center}
    \includegraphics[width=55mm]{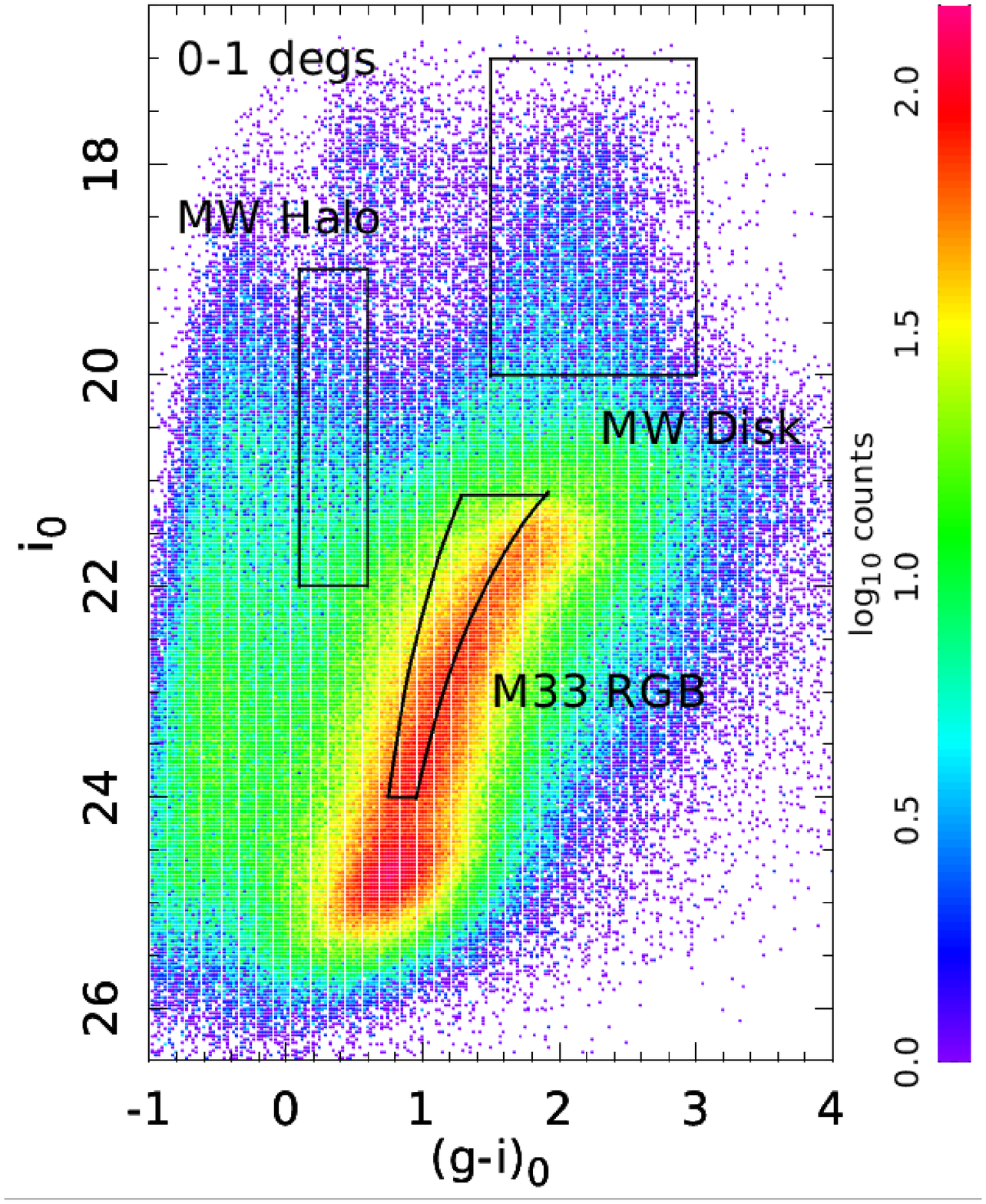}
    \includegraphics[width=55mm]{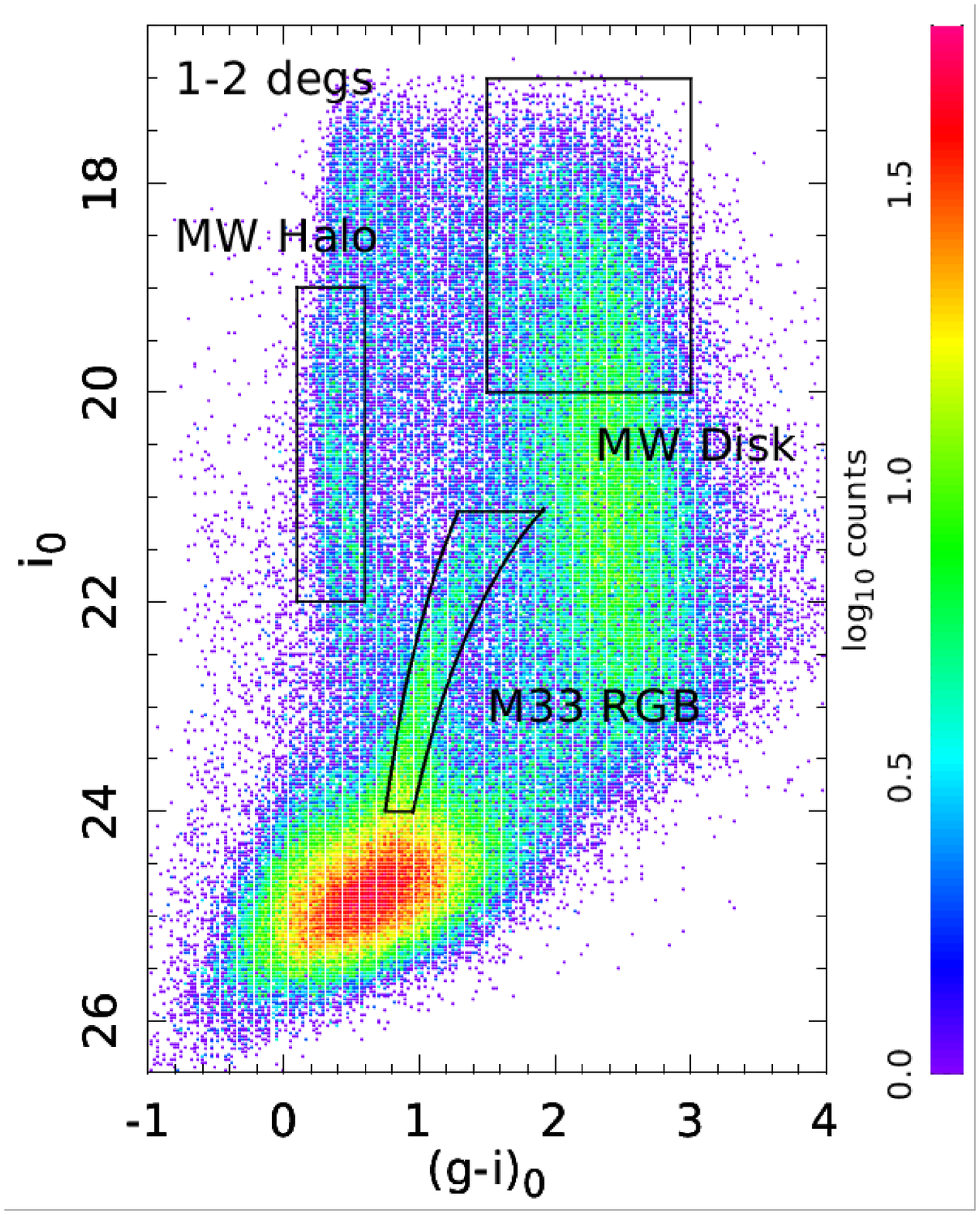}
    \includegraphics[width=55mm]{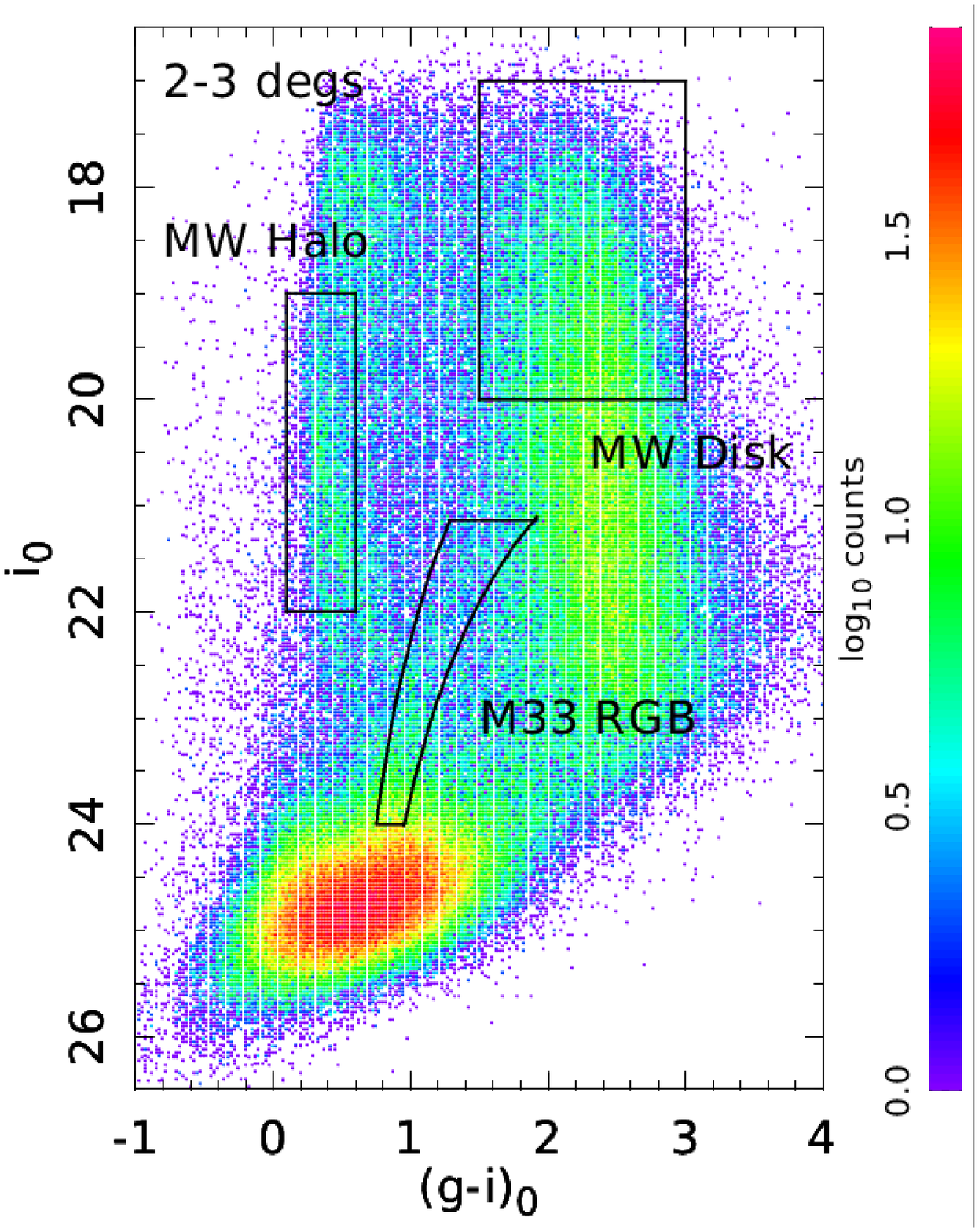}
    \includegraphics[width=55mm]{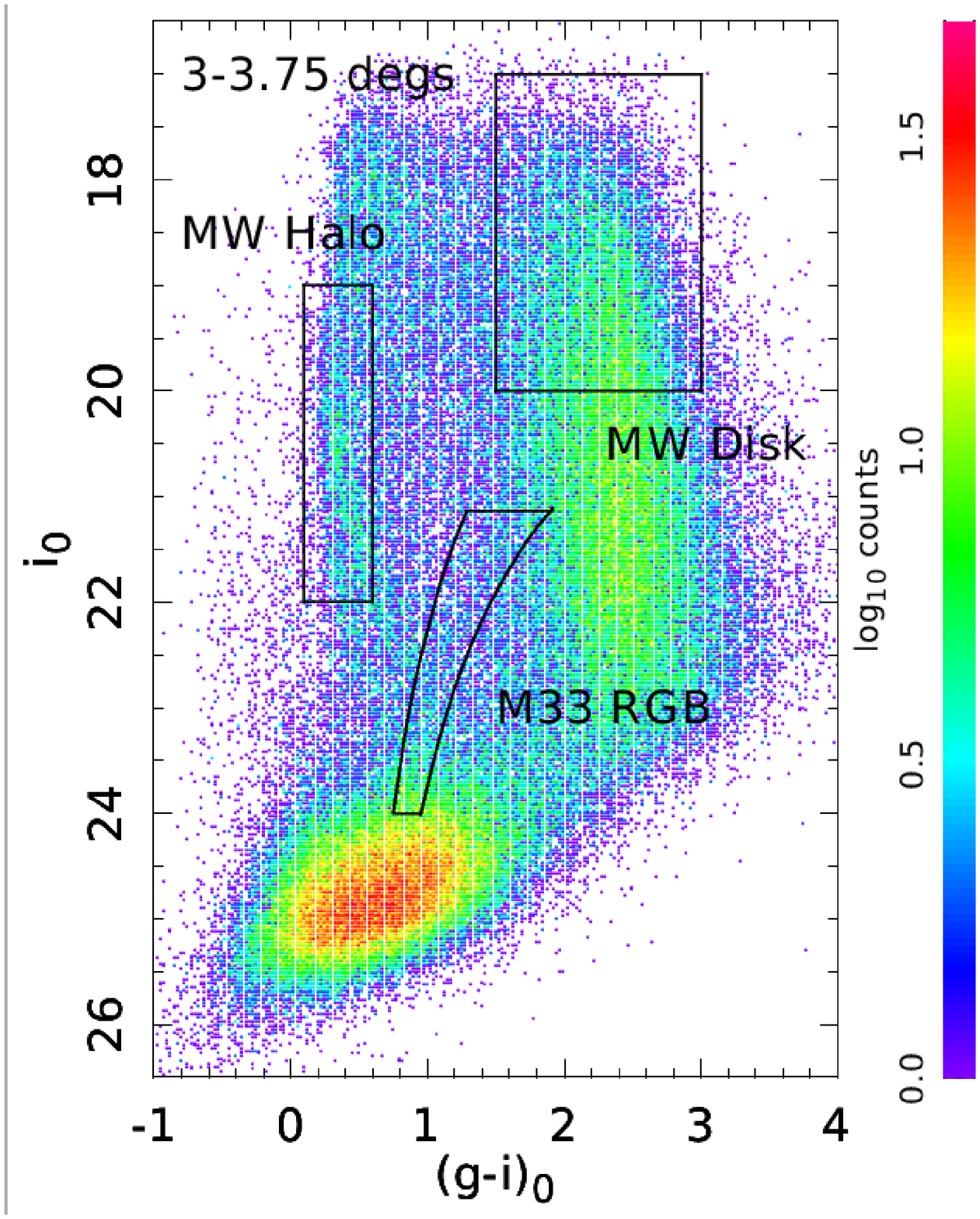}
  \end{center}
  \caption{Color-magnitude (Hess) diagrams of the different annuli
    shown in Figure \ref{frames}.  Bins are 0.025 x 0.025 mag, and are
    shown with a logarithmic scaling in number counts of stars.  Contamination due to the
    foreground MW {\bf halo} and {\bf disk} stars is estimated with the regions
    defined by 0.1 $<(g-i)_{0}<$ 0.6, 19 $<i_{0}<$ 22, and 1.5 $<(g-i)_{0}<$ 3, 17
    $<i_{0}<$ 20,  respectively (shown as the boxes in each
    panel). The isochrones correspond to [Fe/H] = -1.0 and -2.5 dex
    for a 12 Gyr, [$\alpha$/H]=0.0 stellar population at the distance
    of M33, and have magnitude limits of 21.0 $<i_{0}<$ 24.0.  The
    annulus between 3 $< r \le$ 3.75 degrees was used to determine the
    levels of foreground contamination.  The bright clump at $i_{0}
    \sim$ 25, 0 $<i_{0}<$ 1 is mainly composed of misclassified
    background galaxies (with a very small number of M33
    horizontal-branch/red-clump stars).
  }
  \label{cmds}
\end{figure}

Figure \ref{cmds} shows the colour-magnitude (Hess) diagrams for the
data in the annuli in Figure \ref{frames}.  We note that Figure
\ref{cmds}  contains more than 1.4 million
objects that were identified as robust stellar candidates in both $g_{0}$
and $i_{0}$ through the CASU pipeline's object morphological
classification. Magnitudes are de-reddened
source by source using values of $E(B-V)$ in the range 0.034 $\leq
E(B-V) \leq$ 0.130, with $g_{0}$ = $g$ - 3.793 $E(B-V)$ and $i_{0}$ =
$i$ - 2.086 $E(B-V)$ \citep{1998ApJ...500..525S}.  
The data is binned in 0.025 x
0.025 mag bins and is shown with a logarithmic scale for the number
counts of stars.  As mentioned
previously, we want to identify M33 RGB stars but first we need to
estimate the level of contamination.  To examine the MW foreground
contamination, we look
at the two sources of contribution from MW stars easily identifiable
in the CMDs.  The MW halo 
turn-off stars are seen as a thin band on the left of the CMDs, and we
use a region defined as 0.1 $<(g-i)_{0}<$ 0.6, 19 $<i_{0}<$ 22 to
measure their relative numbers in each zone.  The red MW disk dwarfs
are seen as a broader band on the right, 
and we identify them in the region 1.5 $<(g-i)_{0}<$ 3, 17 $<i_{0}<$
20.  Both of the regions for the MW disk and halo stars are consistent
with \cite{2010ApJ...723.1038M}.  Finally, M33 RGB stars are selected
by the colour-magnitude locus where we would expect to find RGB
stars.  This locus is defined using isochrones from the
Dartmouth Stellar Evolution Database 
\citep{2007AJ....134..376D, 2008ApJS..178...89D} which are transformed
to the CFHT photometric 
system \citep{2010ApJ...723.1038M}. These isochrones are between the
12 Gyr [$\alpha$/Fe] = 0.0 isochrones, shifted to 
the M33 distance modulus, with metallicities of -2.5 dex $<[Fe/H]<$ -1
dex.  This is a necessarily broad cut to allow for the possible range of
metallicities that may be present in M33's halo, which we expect to be
predominantly metal poor. Note that
metal-rich stars may also be present, but will likely contribute a
small amount to the overall halo component, while increasing
dramatically the contamination from foreground stars that occupy a
similar locus in the CMD. We could expect some $\alpha$-enhancement in
the M33 halo, as we see in the MW halo (e.g.,
\citealt{2004AJ....128.1177V}), but since there is no evidence to
suggest this we adopt [$\alpha$/Fe] = 0.0 for simplicity. We also note
that the isochrones are being 
used to help define a locus in the CMD, and an absolute interpretation
of the implied metallicities is not intended (for example, there will
also be age degeneracies). A magnitude limit of 21.0
$<i_{0}<$ 24.0 is also imposed on the RGB candidate stars, with the lower 
limit ensuring a high level of completeness while excluding the
majority of bright background galaxies mis-identified as stars (which
becomes a major source of contamination at faint magnitudes; $i_{0}
\approx$ 25, 0 $\lesssim i_{0}\lesssim$ 1).  We test the effect of
raising the faint limit to $i_{0}<$ 23.5 in Section \ref{radial_profile}. 

The four panels in Figure \ref{cmds} correspond to annuli
with the radii between $r$ = 0-1, 1-2, 2-3 and 3-3.75 degrees.  We use
the latter annulus to estimate the spatial variation in the MW
foreground since any M33 halo component, if present, is likely to be
very weak.
The number of stars in each annulus, and the number of stars within
each of the three selection regions, are shown in Table \ref{numbers}.

\subsection{Extended Disk Substructure}

\begin{figure}
  \begin{center}
    \includegraphics[width=110mm,angle=-90]{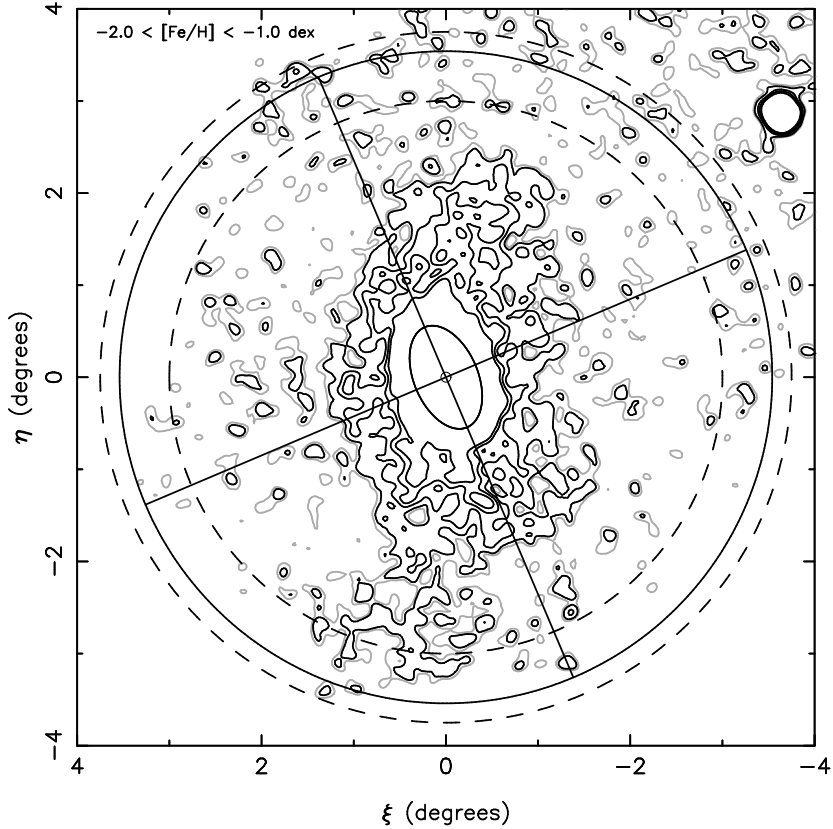}
  \end{center}
  \caption{Density contours of candidate RGB stars similar to Figure
    13 in \cite{2010ApJ...723.1038M} but updated using data from 2010B
    for frames tb62-tb66 (see Figure \ref{frames}).  The grey contour
    is 1$\sigma$ above the background, corresponding to an
    estimated surface brightness limit of $\mu_{V}$ = 33.0 mag
    arcsec$^{-2}$.  We exclude regions within this contour for our
    estimate of the stellar halo.  The black contours correspond to 2,
    5, 8 and 12$\sigma$ above the background ($\mu_{V} \approx$ 32.5, 31.7, 31.2,
    and 30.6 mag arcsec$^{-2}$, respectively).  The feature at ($\xi$,
    $\eta$ = (-3.5, 3.6) is Andromeda II.}
  \label{contours}
\end{figure}

Figure \ref{contours} is a revised
version of Figure 13 in \cite{2010ApJ...723.1038M}, using the new data
in images tb62-tb66 (see Figure \ref{frames}).  The map was created in
an identical way to \citeauthor{2010ApJ...723.1038M} (see their
Section 3.2.2 for details). Figure \ref{contours}
shows the density contours of candidate RGB stars, and uses a slightly narrower
metallicity cut of -2.0 dex $<[Fe/H]<$ 
-1.0 dex than the cut we impose on the candidate RGB stars in the CMDs.  This
narrower cut is used simply because this is the metallicity range in which
the extended disk substructure component is strongest.  There is hardly any
contribution to the extended disk substructure from stars with metallicity between
-2.5 dex $<[Fe/H]<$ -2.0 dex; however, we would not necessarily expect this to be
true of M33's halo RGB stars.  The single grey
contour represents 1$\sigma$ above background, or an estimated surface
brightness limit of $\mu_{V}$ = 33.0 mag arcsec$^{-2}$.  The other
(black) contours are 2, 5, 8 and 12$\sigma$ above the background
($\mu_{V}$ = 32.5, 31.7, 31.2, and 30.6 mag arcsec$^{-2}$,
respectively).  

\begin{figure}
  \begin{center}
    \includegraphics[width=150mm]{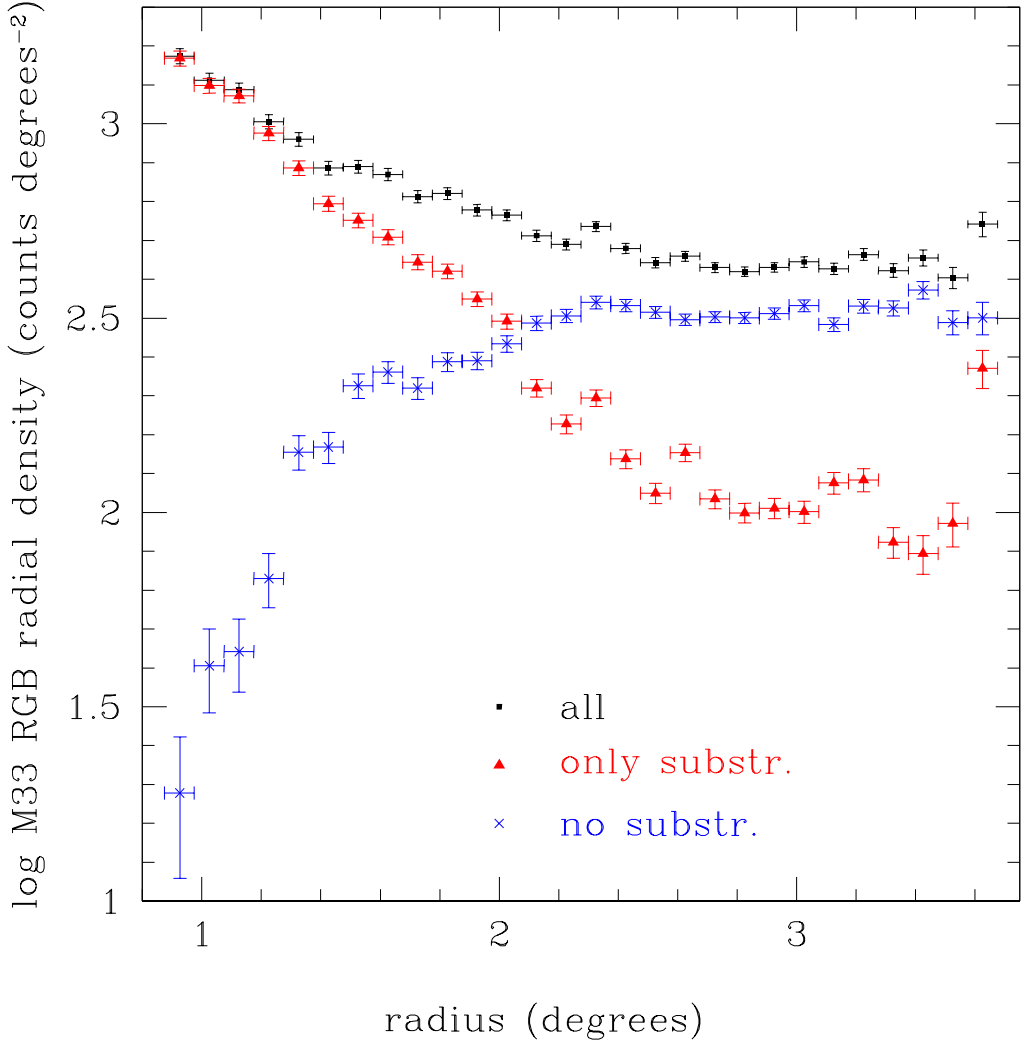}
  \end{center}
  \caption{The radial profiles of extended disk substructure (red triangles) and
    non-substructure (blue crosses) regions, identified by the grey
    1$\sigma$ contours in Figure \ref{contours}, normalized using the
    total annulus area.  The total radial profile is shown by the
    black squares.  We only show the profiles beginning at
    $\sim$0.85, within which the extended disk substructure completely dominates.
    Each bin location is fixed for all three components, and all bins
    have a fixed width of 0.1 degrees (shown by the horizontal error
    bars).  The vertical error bars show $\sqrt{n}$/area.}
  \label{comparing_substr_profiles}
\end{figure}

Figure \ref{comparing_substr_profiles} shows the contributions to the
total radial profile from the regions defined both within and
excluding the 1$\sigma$ contour shown in 
Figure \ref{contours}.  The profiles for the extended disk substructure and
non-substructure regions are normalized using the total annulus area.
The non-substructure regions are seen to start dominating the profile
for $r >$ 2 degrees.  We exclude
data within the 1$\sigma$ contour when probing for the stellar halo.
When we excise the extended disk substructure area denoted by the
1$\sigma$ contour, note that we cannot probe radii smaller than $r
\lesssim$ 1 degree.

\subsection{Foreground and Background Contamination}

\begin{figure}
  \begin{center}
    \includegraphics[width=80mm]{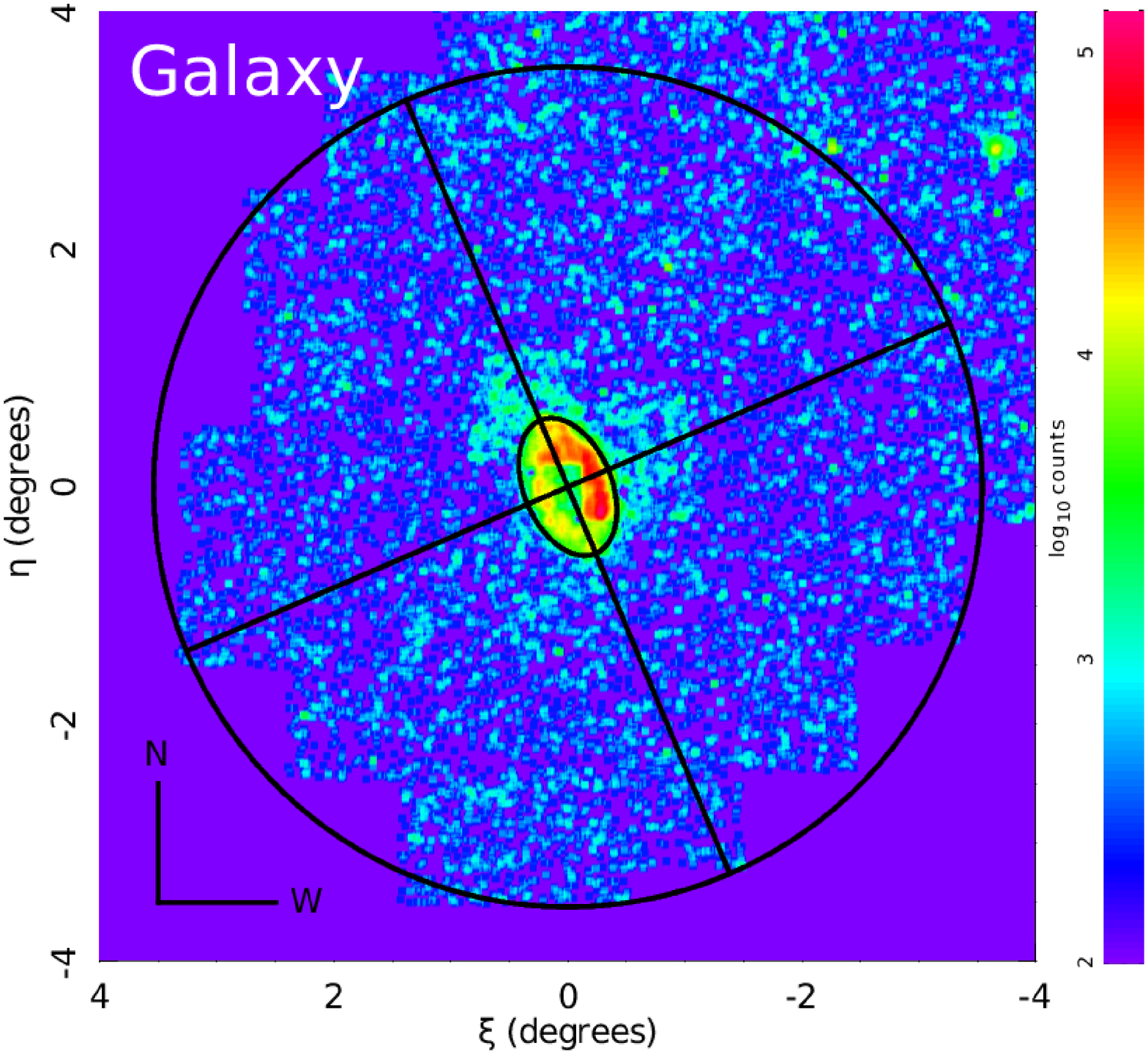}
    \includegraphics[width=80mm]{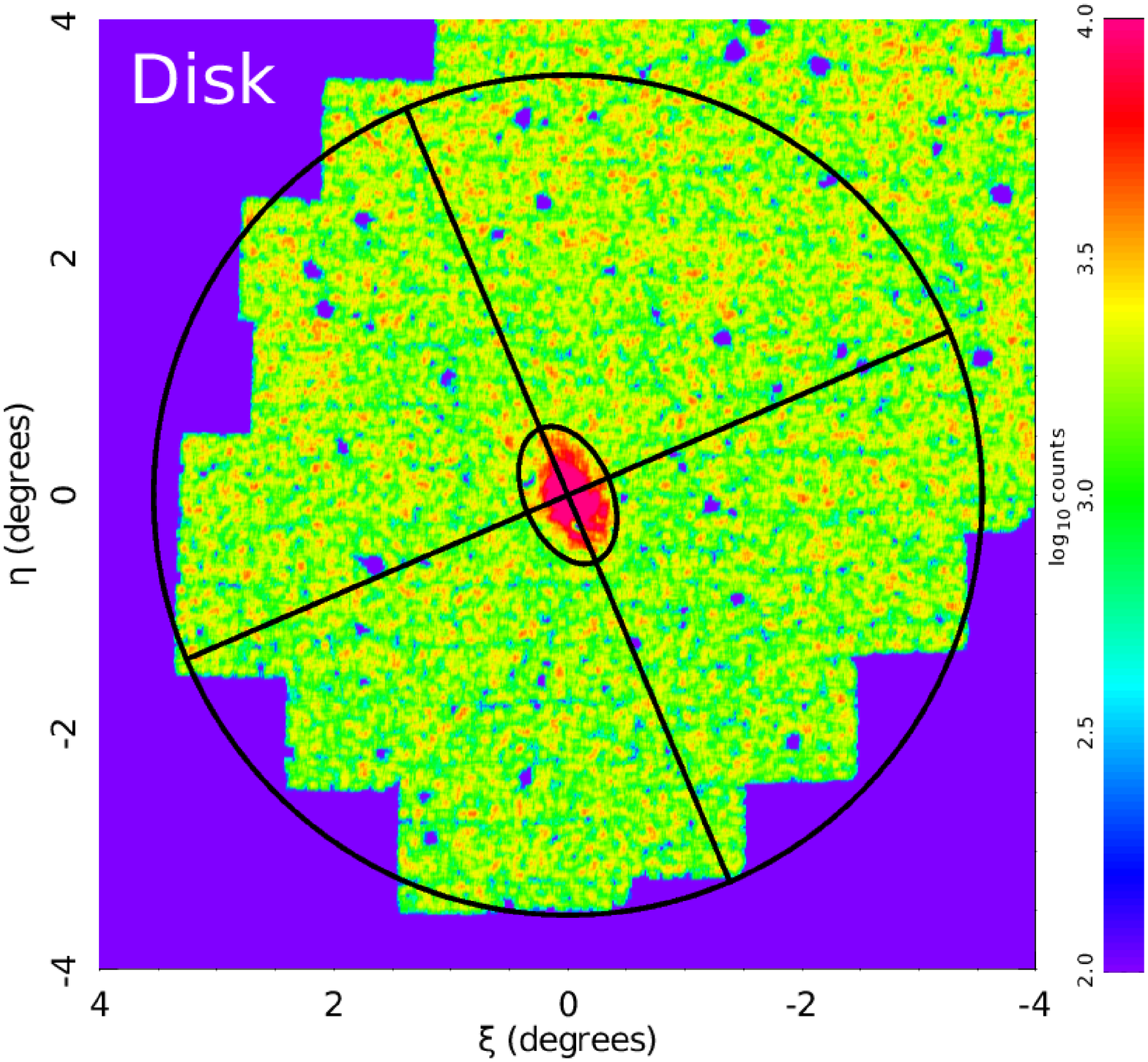}
    \includegraphics[width=80mm]{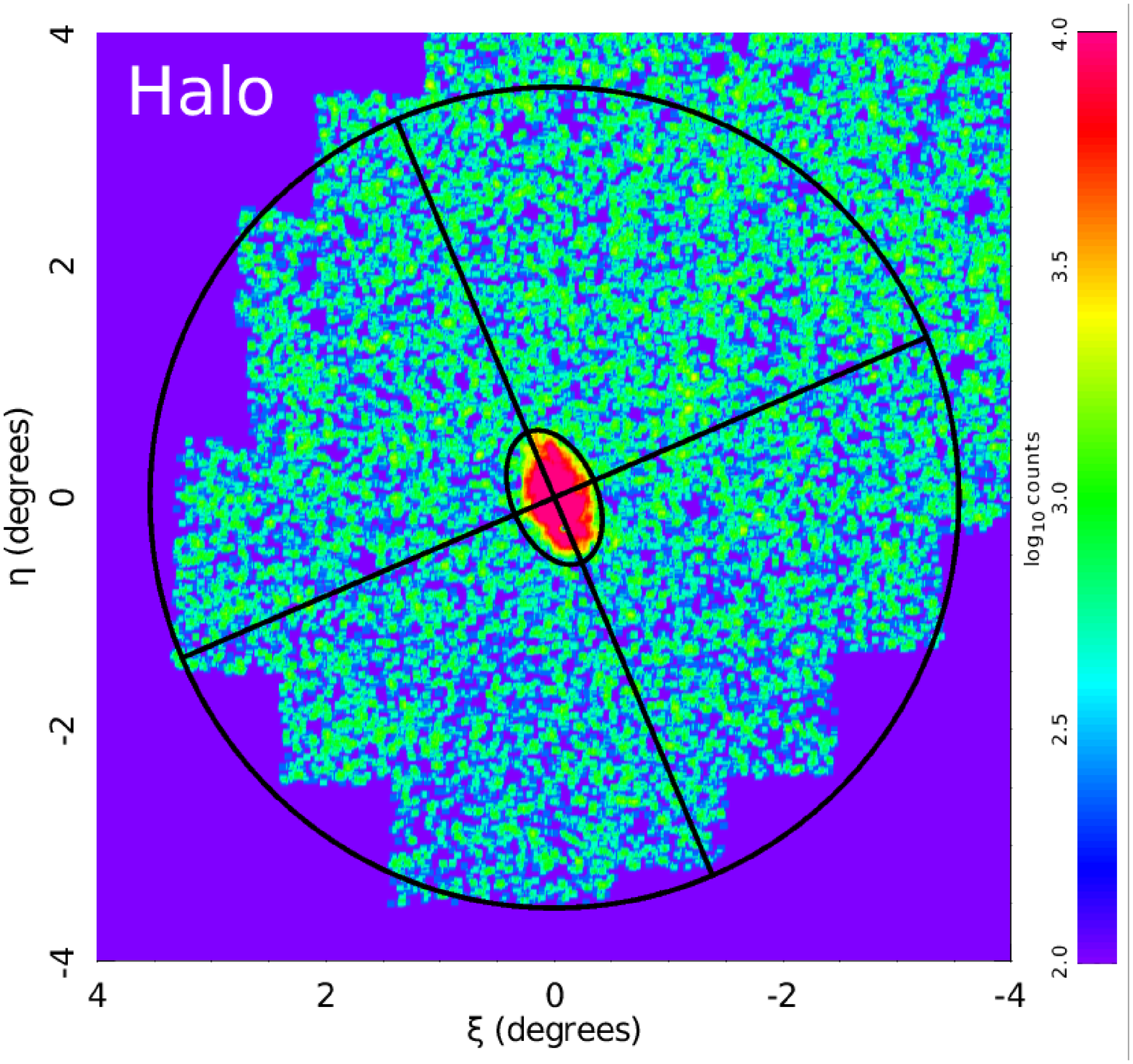}
    \includegraphics[width=80mm]{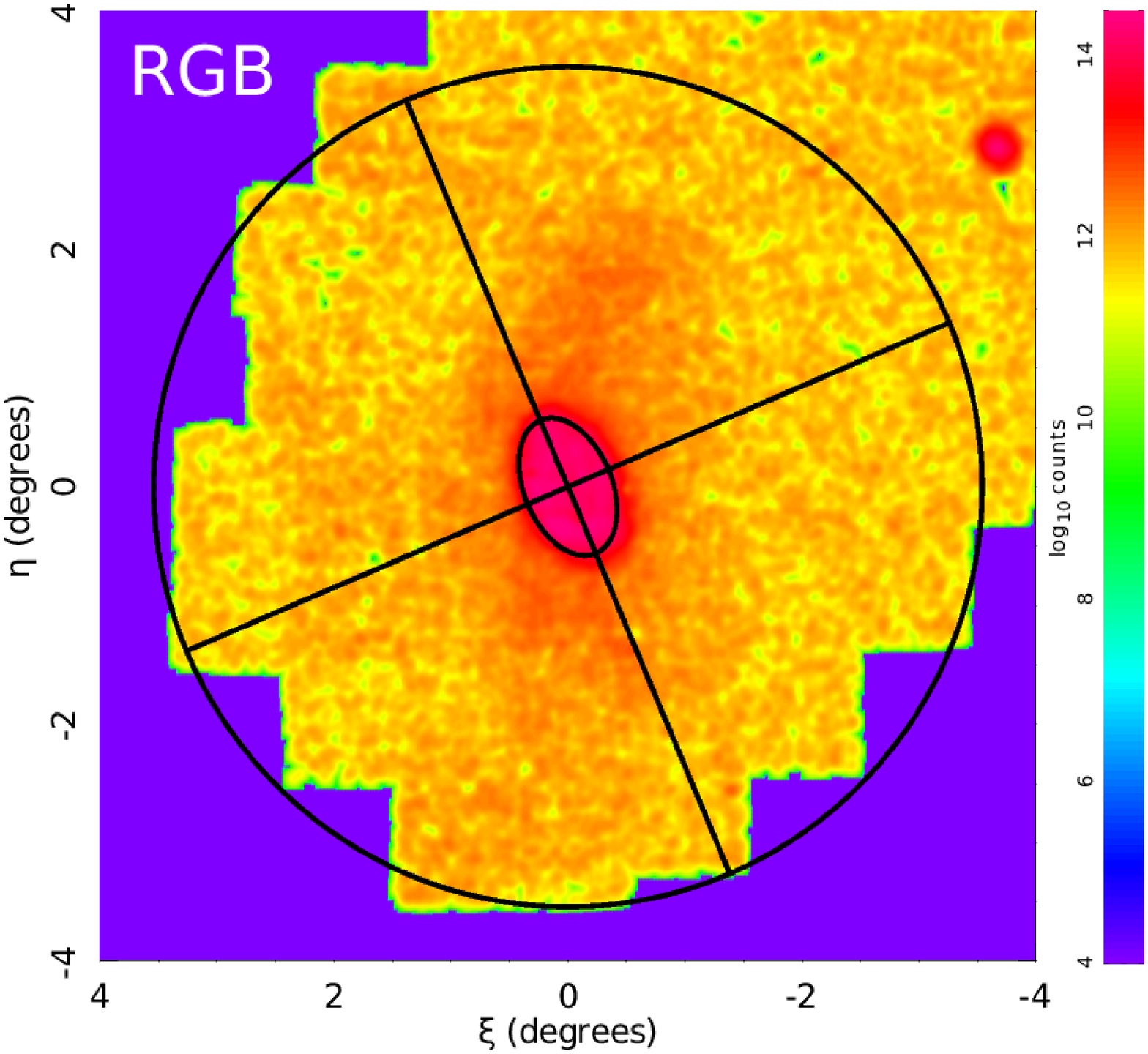}
  \end{center}
  \caption{Smoothed maps of the spatial distribution maps for the
    candidate galaxies, MW disk, MW halo and M33 RGB stars.  See Section
    \ref{methods} for details.  Clearly visible in the RGB map (-2.5 $<$
    [Fe/H] $<$ -1.0 dex) are the M33 extended disk substructure and Andromeda II in the
    NW.  The ellipse, two perpendicular lines and circle are as in
    Figure \ref{frames}.  To ensure the most effective colour range to
    show features (or lack thereof), zeropoints were set at log counts
    = 2 for the galaxy, disk and halo plots, and log counts = 4 for
    the RGB plot. 
  }
  \label{spatial_dist}
\end{figure}

We have identified candidate stars for the M33 RGB, and MW disk and
halo populations, and we have identified the regions
associated with the extended disk substructure surrounding the disk. 
We now test the populations for variations in the spatial
distributions.  Figure \ref{spatial_dist} shows smoothed non-excised
maps of the spatial distribution for the background galaxies and each
of the three populations identified in Figure \ref{cmds} (i.e., the MW
disk, MW halo and the M33 RGB candidate stars).   We identify
background galaxies morphologically using the CASU pipeline, and those
shown in Figure \ref{spatial_dist} have had broad colour and magnitude
cuts applied (17 $< i_0 <$ 23.5, 17 $< g_0 <$ 23.5, and -2 $< (g-i)_0
<$ 4). 

The data is binned into 18 x 18 arcsecond cells.  The galaxy,
disk and halo maps are smoothed once, and the RGB map is  smoothed
three times, all with a boxcar size of 13 x 13 cells (or equivalently
3.9 x 3.9 arcminutes; exactly four times smaller than in
\citealt{2010ApJ...723.1038M}).  The RGB map is smoothed three times
to better highlight the faint extended disk substructure surrounding the disk. 

The galaxy, MW disk and MW halo maps clearly show no significant
global features, although the centre of M33 is apparent due to the
crowded nature in this region where the 
automated object morphological classification is less successful.
Apparent holes in the data are caused by bright foreground stars
preventing detection of faint objects in their surroundings.  The
galaxies misidentified as stars in our sample are expected to have a
similar distribution to the galaxies shown in the galaxy map.  The
RGB map shows the extended substructure surrounding M33's disk, 
and Andromeda II to the north-west. 

Now we excise the regions associated with the extended disk substructure
surrounding the disk and investigate the
variations of the MW disk and MW halo populations in different regions
on the CMD within the 3-3.75 degree annulus (in which we expect little
contribution from bona-fide M33 stars).

\begin{figure}
  \begin{center}
    \includegraphics[width=130mm]{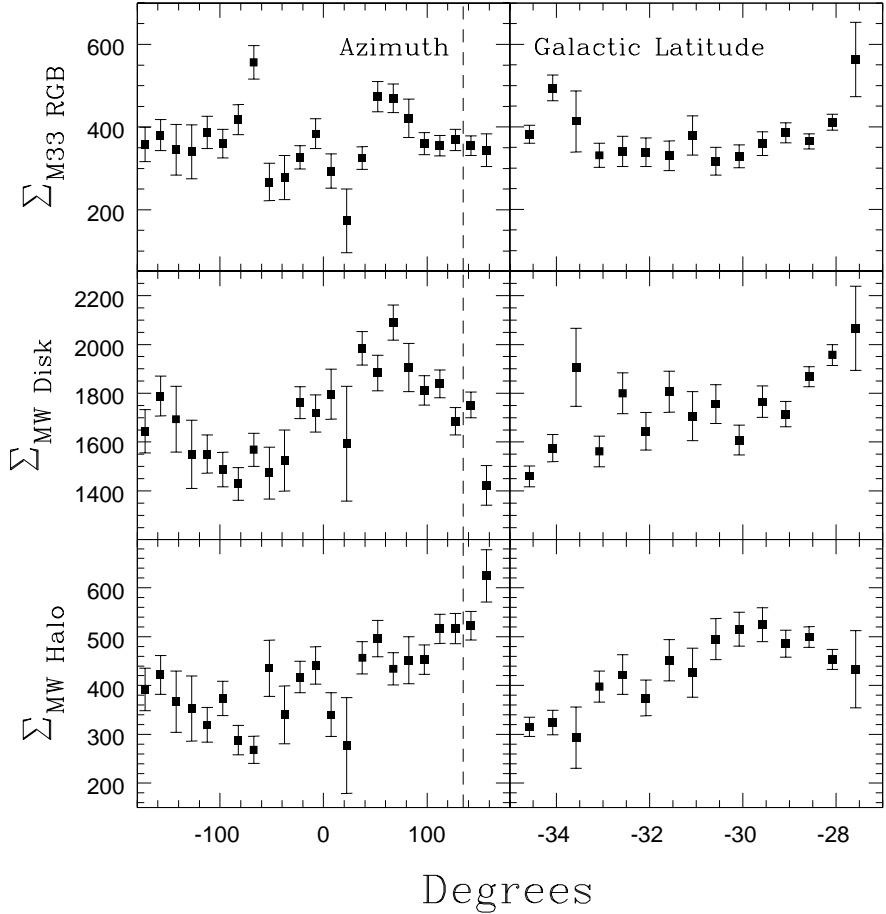}
  \end{center}
  \caption{The left- and right-hand columns show
    the azimuthal and Galactic latitudinal distributions, respectively,
    of the density (counts degree$^{-2}$) variations for each region
    in Figure \ref{cmds}.  The data within the 3 $< r \le$ 3.75 degree
    annulus, having excised the area associated with the extended disk
    substructure, is shown.
    For the azimuthal distributions, 0, 90, $\pm$180, and -90 degrees
    correspond to east, north, west and south, respectively.  M31's
    centre is at approximately 135 degrees in this orientation (as
    indicated by the dashed line).  The errors in all panels correspond
    to the values of $\sqrt{n}$/area.  
  } 
  \label{separate_gal_lat}
\end{figure}

Figure \ref{separate_gal_lat} shows the variation of these three
populations with respect to the azimuthal (left-hand column) and
Galactic latitudinal (right-hand column) distributions.   
All panels show the variation between 3 $< r \le$ 3.75 degrees, with
extended disk substructure regions excised.  Each of the three rows
shows the density  
variation of M33 RGB, MW disk and MW halo candidate stars.  
The density of disk stars increases
towards the disk, as does the density 
of the stars in the MW-halo selection region but with a smaller amplitude.  As we do not expect
the halo stars to vary in latitude in this manner, this suggests some
cross-contamination with the thick disk stars.  

Within the RGB selection shown in Figure \ref{separate_gal_lat}, there
is little variation in the annulus at large radii.  Indeed, the best
fit weighted least-squares fit in both RGB panels is consistent with a
slope of zero.  As such, we conclude that there is no reason to adopt
a spatially-varying foreground for our analysis, and instead use a
constant background, $\Sigma_{bg}$.  

\subsection{Radial Profile}
\label{radial_profile}

Having determined the extended disk substructure area to avoid, we produce
substructure-excised radial profiles.  As previously stated, we expect
M33's halo to be extremely faint and centrally-concentrated so we bin
the data in annuli centred on M33, where we require a certain
signal-to-noise ratio for the bins in each profile.

\begin{figure}
  \begin{center}
    \includegraphics[width=150mm]{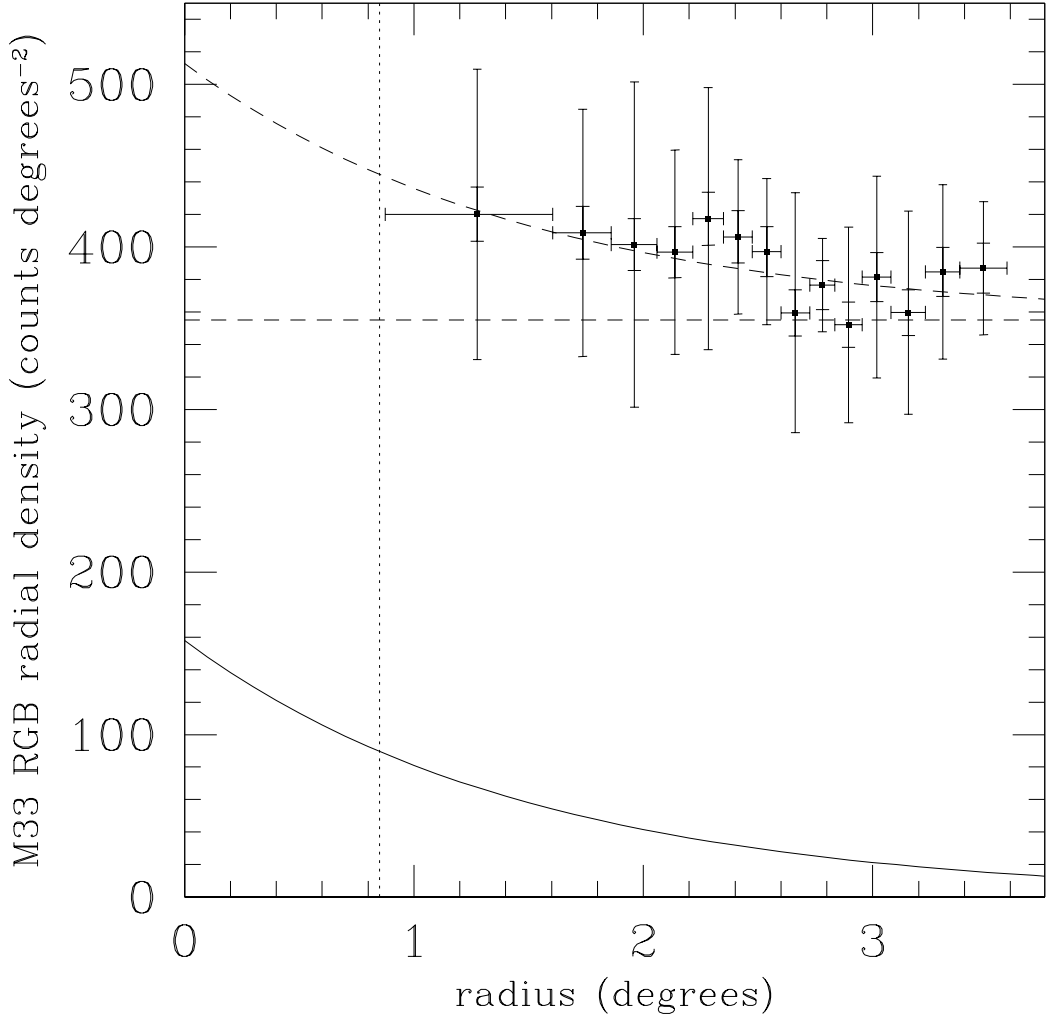}
  \end{center}
  \caption{The background-uncorrected (upper dashed curved line and points) and
    the background-corrected (lower solid curved line) radial profiles
    of RGB candidate stars.  The horizontal dashed line indicates the background
    level, $\Sigma_{bg}$.  Two radial density (vertical)
    error bars are shown: the smaller set is calculated using
    $\sqrt{n}$/area as the error in 
    each bin.  The bin size was allowed to vary until the required
    signal-to-noise ratio of 25 was reached.  Horizontal ``error''
    bars show the width of the bin. The vertical dashed regions
    indicate the radius within which we do not have any data because
    we excise the area dominated by the disk and extended disk
    substructure surrounding the disk.  The larger vertical error bars
    show the variation due to residual substructure (see Section
    \ref{radial_profile} for details).   
  }
  \label{rad_dens_background_uncorrected_corrected}
\end{figure}

Figure \ref{rad_dens_background_uncorrected_corrected} shows radial
density profiles of the RGB stars after excising extended disk substructure
regions.  The small vertical radial density error bars are calculated
using $\sqrt n$/area as the error on the mean of the star counts in
each stellar population. The horizontal error bars indicate  the width
of the annulus.  The size of the annulus was allowed to vary until the
signal-to-noise reached the required value (where the ``noise'' is the
radial density uncertainty).  Each bin in Figure
\ref{rad_dens_background_uncorrected_corrected} has a signal-to-noise
(S/N) cut of 25.  We also use different S/N cuts, but later show that
the results are statistically the same.   The larger error bars shown
in Figure \ref{rad_dens_background_uncorrected_corrected} show the
variation due to residual substructure.  These latter errors were
measured as the standard deviation of number counts between azimuthal
bins (36 degrees in width) around a given radial annulus.   

In all the radial profiles, we see evidence 
for a low-luminosity and centrally concentrated profile in M33's RGB
stars, which is beyond the extended disk
substructure surrounding the disk, and has not previously been seen.
For illustrative purpose only, as this component is so faint and the error
bars are large, we use a Levenberg-Marquardt least squares method to
fit the following exponential model, as shown by the curved lines in Figure
\ref{rad_dens_background_uncorrected_corrected}:  

\begin{equation}
  \Sigma(r) = \Sigma_0 exp\left(-\frac{r}{r_0}\right) + \Sigma_{bg};
\end{equation}

\noindent 
The data points are overlaid with the best fit, as shown by the
curved dashed line.  The horizontal dashed lines show the background
level estimated by the fit.  We also show the background-subtracted fit
with the solid curved lines at the bottom of each panel, where the use
of a constant background is justified in the previous section.  Table
\ref{fit_params} shows the parameters associated with each of the fits
at different S/N cuts, including the S/N = 25 cut 
shown in Figure \ref{rad_dens_background_uncorrected_corrected}.
As previously mentioned, although the parameters vary slightly for
each different S/N cut used, they are statistically the same.

We also test the effect of raising the faint limit of the RGB
selection criteria to $i_{0}=$ 23.5 from $i_{0}=$ 24.0 magnitudes.
When we make the brighter magnitude cut we exclude more contamination -
as we would expect - but we include proportionally less signal. The form
of the radial profile is essentially the same, although less
defined. We therefore continue the analysis using the $i_{0}=$ 24.0
cut.   

We approximate an equivalent surface brightness scale by using the
conversion between stars counts and surface brightness described in
\cite{2010ApJ...723.1038M} (specifically, for $n_{RGB} <$ 350 stars
degree$^{-2}$ from their Figure 15).  For details of this
conversion, see \cite{2010ApJ...723.1038M}, but note that {\it the
conversion is only an approximation}. The RGB stars in
\citeauthor{2010ApJ...723.1038M} are selected using -2 $<$ [Fe/H] $<$
-1 dex, and $i_0 <$ 23.5 magnitudes, whereas here we use -2.5 $<$ [Fe/H] $<$
-1 dex, and $i_0 <$ 24.0 magnitudes.  There are also large
systematic uncertainties inherent in the technique. 

We estimate the luminosity of this component by first
simply summing the total number of stars contributing to the profile
in the radial range for which we have data (i.e., 0.88 $\leq  r \leq$
3.75 degrees), and using the conversion as above.  Assuming Poisson
statistics, we obtain 765 $\pm$ 95 stars (assuming a background of
355 stars degree$^{-2}$, and without propagating the uncertainty in
the background), corresponding to a luminosity of $L$ = 2.4 $\pm$ 0.4 x
10$^{6} L_{\odot}$.  Note that this initial estimate is independent of
any assumptions we could make about the profile of the component.

\begin{figure}
  \begin{center}
    \includegraphics[width=150mm]{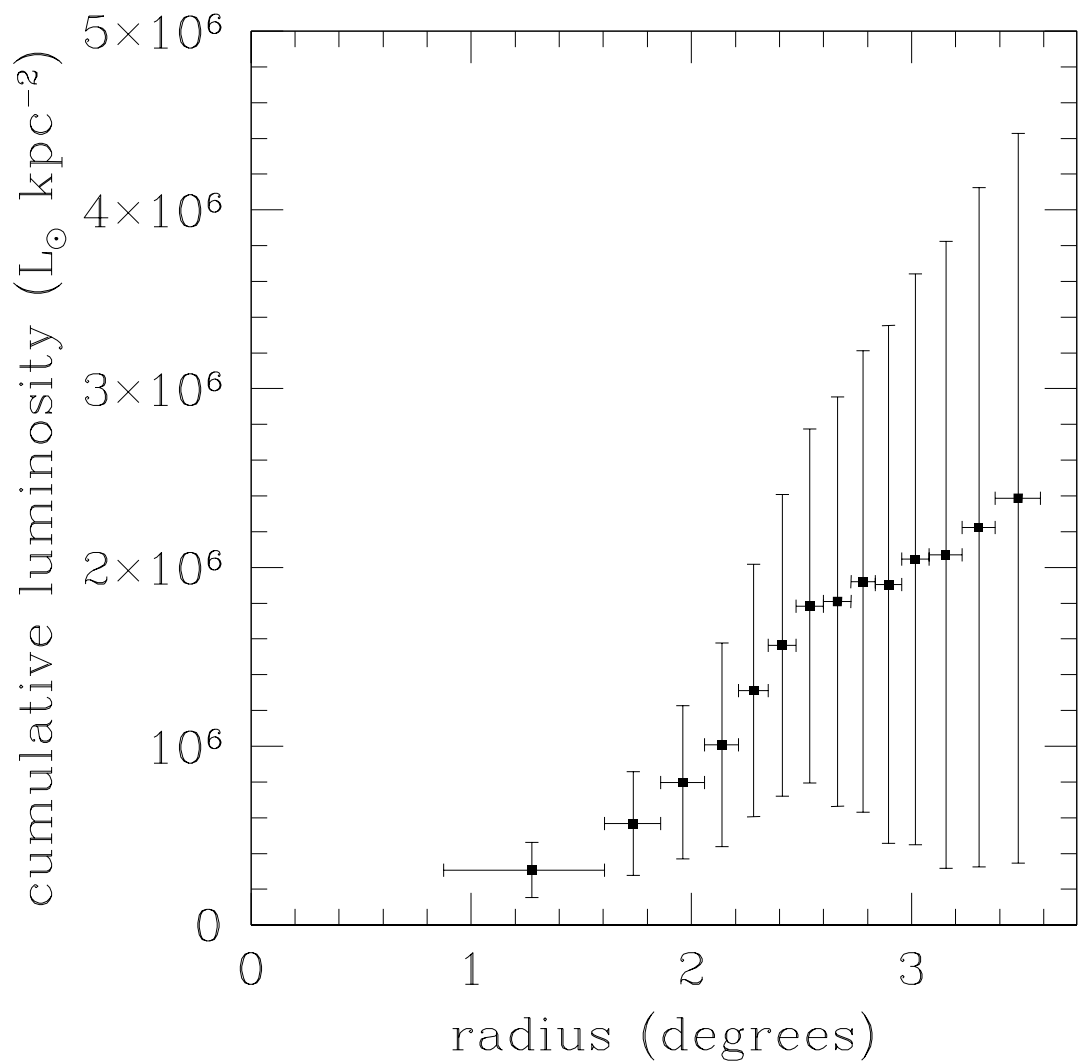}
  \end{center}
  \caption{The cumulative luminosity as a function of radius.  Horizontal
    ``error'' bars show the width of the bin, and are the same as
    those shown in Figure
    \ref{rad_dens_background_uncorrected_corrected}.  The error on the
    luminosity is calculated by combining the Poisson errors for the
    RGB candidate star counts with the uncertainty of the background.
    This figure highlights the increase in our estimate of the
    luminosity uncertainty as we increase the area we consider.
  }
  \label{cumulative_luminosity}
\end{figure}

To calculate the luminosity extrapolated to the center, however, we
assume a spherically symmetric smooth profile that is described by the
exponential fit with a scale length of 1.5 degrees (or 21.1 kpc). We
calculate the fraction of
the integral of the exponential fit between 0.88 $\leq  r \leq$
3.75 degrees compared to 0 $\leq  r \leq$ 0.88 degrees, allowing us to
calculate the luminosity between 0 $\leq  r \leq$ 3.75 degrees.  This
simple extrapolation yields an estimate of $L$ = 3.8 $\pm$ 0.5 x
10$^{6} L_{\odot}$.  (If we use a similar technique to extrapolate
under the whole exponential curve, i.e., out to infintiy, we obtain
$L$ = 4.1 $\pm$ 0.5 x 10$^{6} L_{\odot}$.)  We note that the large 
uncertainties on the exponential profile fit to our data and the unknown
intrinsic profile of this component, make these extrapolated estimates
highly uncertain.  As noted above, we also do not include the error on
the background. The effect of including this is seen in Figure
\ref{cumulative_luminosity}; as we integrate out to larger radii, the
relative luminosity error estimates increase.  At larger radii, there
are fewer candidate RGB stars but a relatively larger contribution
from the background.


\section{Discussion and Conclusions}
\label{conclusions}

It was expected that any stellar halo signal around M33 would be at
least as faint as the recently discovered extended optical disk
substructure surrounding the disk \citep{2009Natur.461...66M,
  2010ApJ...723.1038M}.  Hints of a radial fall off beyond the extent
of the extended disk substructure suggested a tentative halo detection
\citep{2010ApJ...723.1038M}. We followed up this possibility in our
present study, using higher spatial resolution maps, excising any
contribution from the extended disk substructure, and subtracting off
contamination from foreground and background sources, so that we are
more able to cleanly resolve and identify any remaining signal.   

We detect a radial density drop off that we interpret as an upper
limit of the M33 candidate stellar halo.  The signal is extremely
faint, but seems robust to various signal-to-noise cuts; as previously
noted, we observe only 765 $\pm$ 95 excess stars between 0.88 $<r<$ 3.75
degrees. 

\begin{figure}
  \begin{center}
    \includegraphics[width=120mm]{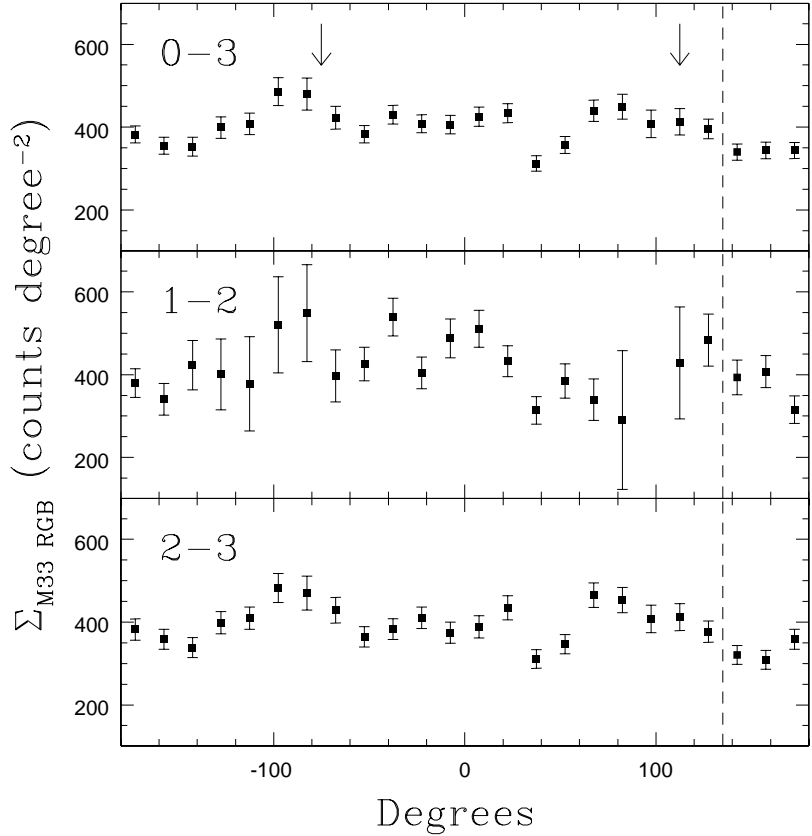}
  \end{center}
  \caption{The azimuthal distribution of RGB candidate stars for the
    annuli shown in the top left of each plot.  0, 90, $\pm$180, and -90 degrees
    correspond to east, north, west and    south, respectively.  M31's
    centre is at approximately 135 degrees in this orientation (as
    indicated by the dashed line).  The top, middle and bottom panels
    show the data with $r <$ 3, 1 $<$ $r$ $<$ 2, and 2 $<$ $r$ $<$ 3
    degrees, with the regions associated with the extended disk
    substructure excised in all panels. The left- and right-hand
    arrows correspond approximately to the SE and NW tips of the
    S-shaped warp of the extended disk substructure.
  }
  \label{rgb_azim}
\end{figure}

Are we justified in claiming this extra component is a halo?
We have azimuthally averaged annuli centred on M33 to find a
low-luminosity and centrally concentrated profile. 
The top panel of Figure \ref{rgb_azim} shows
the azimuthal distribution of the RGB candidate star density within 3
degrees having excised the extended disk substructure regions.  We see contamination of the 
MW foreground stars does not appear to affect the density variation of
these RGB candidate stars (i.e., we do not see a reflection of the
density profiles shown in Figure \ref{separate_gal_lat} for the MW
disk candidate stars). If our extra component was actually
residual low-level emission from the already known extended disk substructure we
would expect to see this reflected in this plot, with overdensities
around the regions associated with the tips of the S-shaped warp,
indicated by the two arrows in the top panel.  The azimuthal
distribution is fairly flat, but we note that overdensities are
apparent near the warp's tips suggesting some contamination from the
extended disk substructure.  We further split the data into two annuli, 1-2
and 2-3 degrees, but we do not see evidence that the RGB candidates show
any major differences in their azimuthal distribution from one another
in either annulus.  

\begin{figure}
  \begin{center}
    \includegraphics[width=170mm]{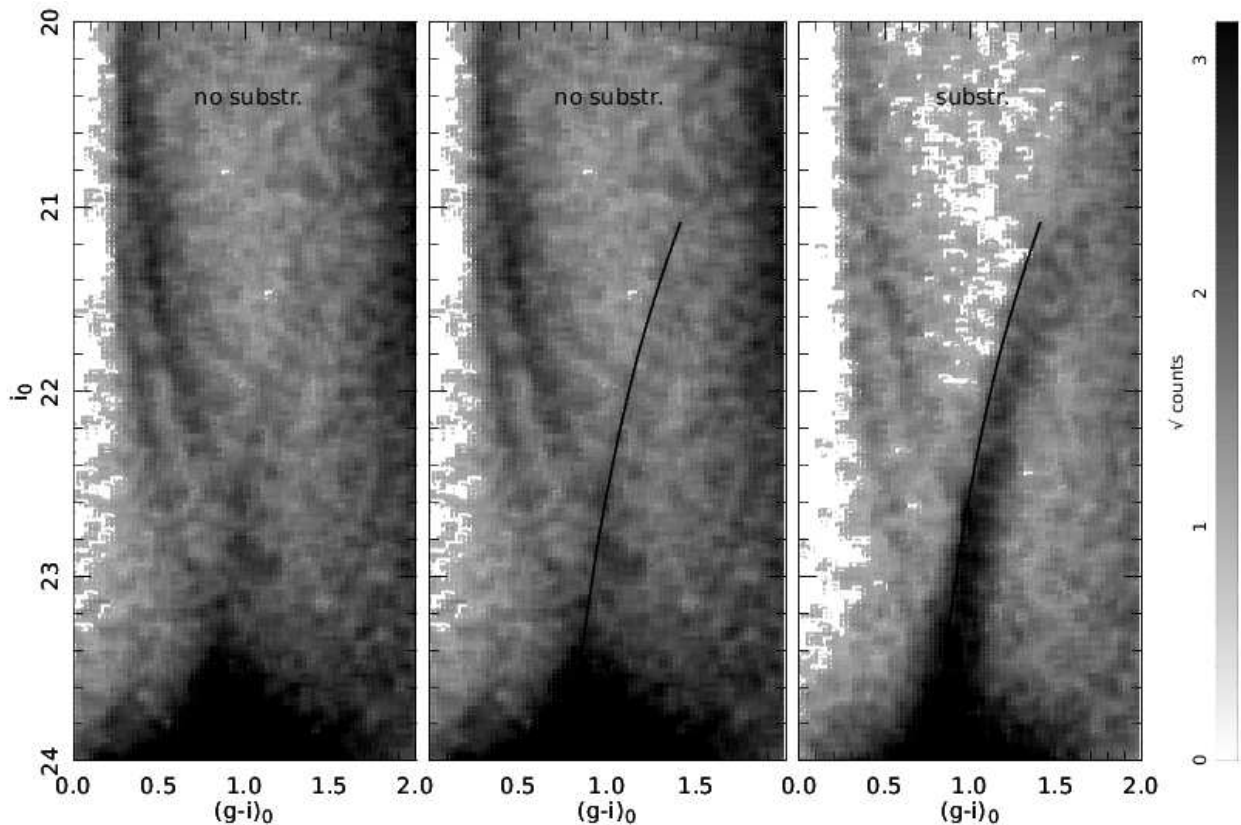}
  \end{center}
  \caption{Similar to Figure \ref{cmds}, but here showing the CMD for
    the region with $r <$ 3 degrees.  The left-hand and middle panels
    show the CMD after excising the extended disk substructure
    areas in Figure \ref{contours} (which effectively imposes a
    minimum radius of $r$ = 0.88 degrees).  The RGB that we aim to detect is
    so faint that it is barely visible on the left hand plot.  We
    overlay a [Fe/H] = -2 dex isochrone on the middle plot.  The
    right-hand plot shows the CMD of the extended disk substructure
    areas for comparison. 
  }
  \label{cmd_0_3degs_no_substr}
\end{figure}

Further constraining this newly discovered component, we show in Figure
\ref{cmd_0_3degs_no_substr} the CMD for all 
objects with  $r < $ 3 degrees, except for those within the extended
disk substructure's 1$\sigma$ contours in Figure \ref{contours} (again,
this imposes a minimum radius of $\sim$1 degree).  We note that the RGB stars 
that we aim to detect are just visible to the eye on the left-hand
plot.  Again, we see the extreme relative faintness of this component.  On
the middle panel we overlay an [Fe/H] = -2 dex isochrone to this
feature.  As expected if this component is a halo, this crude
measurement indicates that it is relatively metal-poor.  The extended
disk substructure metallicity for comparison is [Fe/H] = -1.6 dex
\citep{2010ApJ...723.1038M}.  We show the CMD for the extended disk
substructure in the right-hand panel, again overlaying an [Fe/H] = -2
dex isochrone for comparison.  We can see that the candidate halo lies
on the metal-poor side of the extended disk substructure RGB.  

\begin{figure}
  \begin{center}
    \includegraphics[width=160mm]{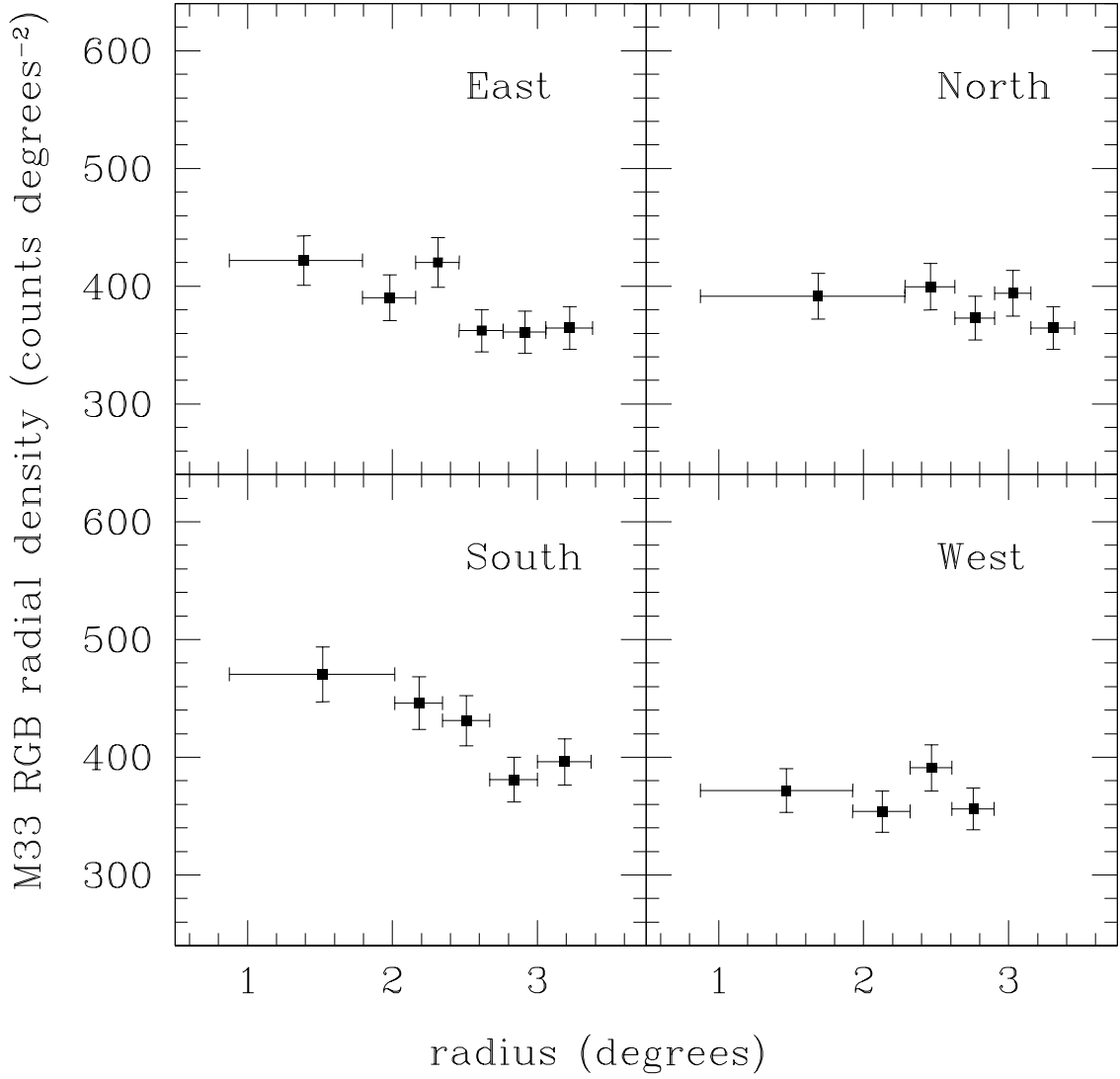}
  \end{center}
  \caption{Background-uncorrected profiles for the quadrants split by
    major and minor axes, e.g., as shown in Figure \ref{contours}.}
  \label{tblr}
\end{figure}

\begin{figure}
  \begin{center}
    \includegraphics[width=160mm]{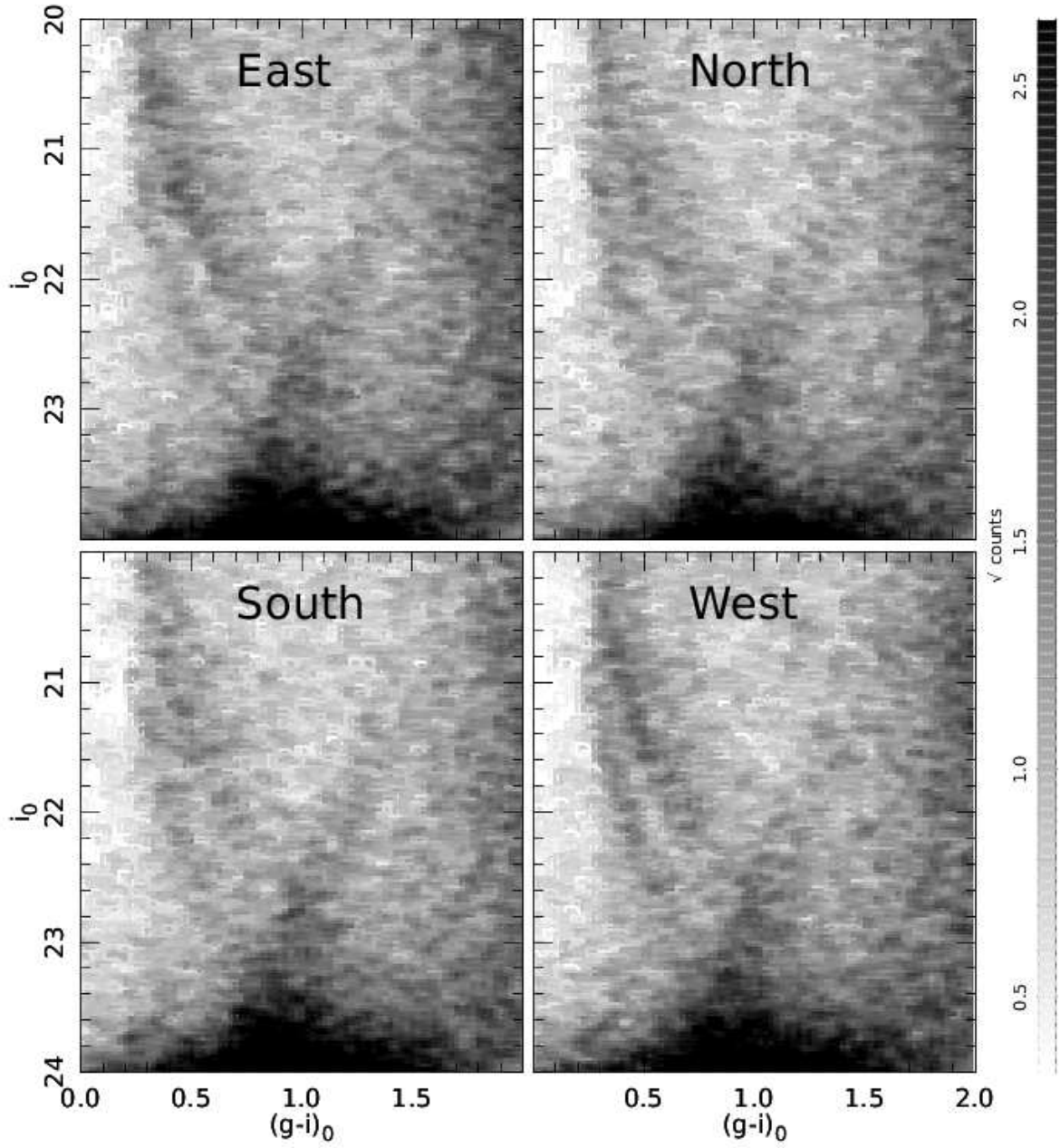}
  \end{center}
  \caption{The CMDs for the quadrants used in Figure \ref{tblr}.}
  \label{tblr_cmds}
\end{figure}

With this component that we identify as M33's candidate halo, it is
appropriate to ask - even with such poor signal-to-noise - if we see
any azimuthal asymmetry.
To test for this, we split the data into the four quadrants split by
the major and minor axes, e.g., as shown in Figure \ref{contours}.
The resulting radial profiles for each quadrant are shown in Figure
\ref{tblr}, and the associated CMDs are shown in Figure
\ref{tblr_cmds}.   It appears as if the east and south quadrants have
the steepest radial declines, whereas the north and west quadrants are
flatter.  In other words, there may be a variation in the radial
profile of the candidate halo when split along the major axis.
Further interpretation of this possible asymmetry must wait for higher
quality and deeper data.

In summary, we have a weak detection that is not clearly indistinguishable in
either azimuthal distribution or metallicity from the extended
disk substructure component.  If this is a halo, we use the estimates so
far obtained to place upper limits on the luminosity.  We note that
with the data at hand we must leave open that it could be another
component, such as a very extended thick disk.

The location of the extended disk substructure was the most important
knowledge prior to beginning this study, in a similar way that the
spectroscopic knowledge of the metal-rich component led to the
discovery of the metal-poor halo in M31 \citep{2006ApJ...653..255C,
  2006ApJ...648..389K}. 
If we directly compare M31's halo with M33's candidate halo
we find that apart from the obvious difference in luminosity
($L_{M31,halo, V}\sim10^{9} L_{\odot}$, \citealt{2007ApJ...671.1591I};
$L_{M33, halo, V} $ = 4.1 $\pm$ 0.5 x 10$^{6} L_{\odot}$.), expected
because of the mass 
difference between the two galaxies, it is unclear if the exponential
scale lengths are significantly different; we estimate M33's scale
length to be 21 $\pm$ 18 kpc, similar to that found for M31  
($\sim14$ kpc; \citealt{2005ApJ...628L.105I}).  In light of the
discovery here, the spectroscopic work by \cite{2006ApJ...647L..25M}
needs to be revisited so that a comparison of M31 and M33's halo
metallicity can be made (see \citealt{Trethewey2011thesis}, and a
forthcoming paper by \citealt{Trethewey2012_in_prep}).  

As mentioned in Section \ref{m33}, the favoured interpretation to
explain M33's extended disk substructure surrounding the disk is a tidal
interaction with M31 \citep{2009Natur.461...66M, 2010ApJ...723.1038M}.
It is extremely likely that this interaction 
also affected M33's halo: at least altering if not stripping it, with
some of M33's halo then being accreted onto M31. The halo could also
extend beyond the point to which the PAndAS data set is able to
measure it ($\sim$ 3.75 
degrees, or $\sim$ 5 degrees to the north-east.)  In the supplementary
material movie of \cite{2009Natur.461...66M}, the 
end of the modelled interaction also includes the stellar halo.  Though a
significant amount of halo material is stripped from M33, appearing to
extend beyond the virial radius to form a low-luminosity bridge
between M33 and M31, most of the 
halo appears to remain bound to M33.  The material stripped from M33's halo
also extends well beyond the area observed in PAndAS.  More of the
remaining bound material appears on the south-east side (away from
M31) than in the north-west (closest to M31).  Our observations appear
to broadly agree with this model, as we see more of a gradient in the
radial profile in the south quadrant.  

The most probable scenario(s) for how M33's candidate halo was built could
be quite different from Milky Way and M31 because it is approximately
ten times less massive than either.  Unlike M31, M33 has no bulge,
 a warped extended disk substructure, and is likely interacting with a much
 more massive neighbour. Studies of the M33 outer halo clusters also
 suggest that with the low GC surface density ($\Sigma_{GC, M33} \sim
 0.14$ deg$^{-2}$) compared to M31 ($\Sigma_{GC, M31} \sim 0.8$
 deg$^{-2}$), M33 either had a much calmer accretion history than M31
 or that some of the outer halo clusters could have been tidally
 stripped by M31 \citep{2009ApJ...698L..77H, 2011ApJ...730..112C}.
 The latter idea obviously supports the favoured interpretation that
 could explain the warped extended disk substructure component.

We now compare our results with a model that predicts the size of
M33's stellar halo.  \cite{2007ApJ...666...20P} use an analytic model with empirical
constraints from $z \sim$ 0 observations to predict the fraction of
stellar halo mass compared to the total luminous mass.  They define
the diffuse stellar mass fraction as $f_{IHL} =
M^{diff}_{*}/M^{total}_{*}$; note that they use mass rather than
luminosities, to avoid uncertainties involved with luminosity
evolution. There is no 
distinction made between the substructure in the halo, or the smooth
diffuse halo that might underlie the substructure.  Their predictions
cover a range of host galaxy's dark matter halo masses, from small
late-type galaxies to large galaxy clusters ($\sim 10^{11}$ to $\sim
10^{15}$ $M_{\odot}$).   The stellar material is assumed to be able to
become part of the diffuse stellar halo when its dark matter subhalo
has become significantly stripped.   The dark matter subhalo is considered
disrupted when its maximum circular velocity falls below a critical
value - which is set by considering the empirical constraints. For DM
halos of mass $\sim10^{11} M_{\odot}$ ($\sim M_{M33}$;
\citealt{2000MNRAS.311..441C}), the stellar halo  
luminosity fraction  is expected to be $\le1.0\%$ (thus, $L_{stellar
  halo} \le$ 10$^{7} L_{\odot}$).  A galaxy's mass-to-light ratio (over the
entire halo out to the virial radius) varies as a function of DM halo
mass, and this drives the fraction of stellar halo material; small
galaxies are  expected to accrete material from dwarf galaxies, which
have high mass-to-light ratios and therefore share little luminous
material with their host galaxy's stellar halo
\citep{2007ApJ...666...20P}. Even if they share all of their material,
the contribution is not large. This picture seems to broadly agree with our
upper limit estimate of M33's extremely faint candidate stellar halo, 
$L_{M33, halo, V} $ = 4.1 $\pm$ 0.5 x 10$^{6} L_{\odot}$ (0.4$\% \lesssim
L_{M33, host, V} \lesssim$ 0.5$\%$). 

\begin{figure}
  \begin{center}
    \includegraphics[width=80mm]{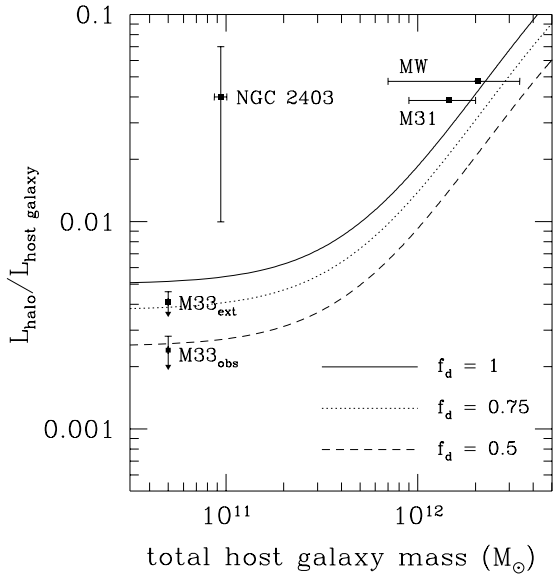}
  \end{center}
  \caption{The halo luminosity as a fraction of
    the host galaxy luminosity against the total host galaxy mass
    (dark plus luminous) for the MW, M31, M33 and NGC 2403.  Data for
    the MW, M31 and NGC 2403 are from the literature (see Sections
    \ref{mw}, \ref{m31} and \ref{conclusions}, respectively, for details and 
    references).  M33's host galaxy mass is from
    \cite{2000MNRAS.311..441C}.  The ranges of M33's halo luminosity
    are our estimates from this paper.  The observed range comes from
    strictly limiting 
    the integration between the range of our actual data, whereas the
    extrapolated range comes from extrapolating inwardly to the centre of
    M33 and outwardly to M33's virial radius.  The lines represent the
    models for the intrahalo light fraction in
    \cite{2007ApJ...666...20P}.  We use $n_{eff}$ = 1 in conjunction
    with the three values for $f_d$ shown in the plot. $n_{eff}$
    represents the effective number of satellites with mass $M_{sat} =
    M_{host}/20$; $f_d$ represents the total stellar mass fraction a
    satellite contributes to its host galaxy
    halo. No distinction is made between halo substructure and a smooth halo 
    component for the host luminosity estimates in the literature for
    the MW, M31, and NGC 2403 - similarly for the
    \citeauthor{2007ApJ...666...20P} models.
    \cite{2010ApJ...723.1038M} estimate the luminosity of the extended
    disk substructure (EDS), $L_{EDS} \approx 0.01 L_{M33}$ so
    $L_{EDS} \approx 10^7 L_{\odot}$.M33's extended disk substructure
    is in the halo region, and if it was to be included in the
    estimates shown in this figure it would raise each
    $L_{halo}/L_{host galaxy}$ estimate by 0.01.
  }
  \label{halo_lums}
\end{figure}

Figure \ref{halo_lums} plots the fraction of halo luminosity compared
to the host galaxy luminosity, against the host galaxy mass for
several galaxies including M33.  The lines represent equation 7 from
\cite{2007ApJ...666...20P} for the model of the intrahalo light, with
different values for the parameters $n_{eff}$ and $f_d$.  Here, $n_{eff}$,
expected to be of order unity, represents the effective number of 
satellites with mass $M_{sat} = M_{host}/20$; $f_d$ represents the total
stellar mass fraction a satellite contributes to its host galaxy halo.  We
note that if we swap the values for $f_d$ and $n_{eff}$, the lines
would vary in the same way.  

The two estimates for M33's candidate halo luminosity fraction
are for the directly observed estimate (0.88 $< r <$ 3.75 degrees),
and the implied, extrapolated estimate ($r <$ 10.64 degrees).  The
values for the MW and M31 are taken from Sections \ref{mw} and
\ref{m31}, respectively.  We also include estimates of the NGC 2403's extended
component \citep{2012MNRAS.419.1489B}, as it has a similar total
stellar mass to M33 (9.4 $\pm$ 0.7 x $10^{10} M_{\odot}$;
\citealt{2002AJ....123.3124F}).   It is at a distance of 3.1 Mpc
\citep{2001ApJ...553...47F}, has an inclination of 63 degrees
\citep{2002AJ....123.3124F}, and is the brightest member of a loose
galaxy group and is therefore considered much more isolated than M33
(the closest large galaxy is M81, which is four times further from NGC
2403 than M33 is from M31;
\citealt{2012MNRAS.419.1489B}). \citeauthor{2012MNRAS.419.1489B} use
Subaru/Suprime-Cam to obtain images 39 x 48 kpc around the centre of
NGC 2403, and see an extended component which could be disk structure
or a halo.   Extrapolating out to 50 kpc they find that the haloes
contain $\sim$ 1-7 $\%$ of the total V-band luminosity, or $L_{2403,halo,V} \sim$
1-7 x 10$^{8} L_{\odot}$, depending on whether or not an exponential or
Hernquist profile is used (if they extrapolate out to 100 kpc the
estimate does not significantly change).

The values of $L_{halo}/L_{host galaxy}$ for the MW and M31 are close
to those of the models by \cite{2007ApJ...666...20P}.  However, for
the less massive galaxy NGC 2403 we find the models seem to
underestimate the contribution of halo light from these smaller
galaxies - and may also do the same for M33, although with such a weak
signal as we detect here there are large uncertainties.  If we further
include M33's extended disk substructure ($L_{substructure}$ $\sim$ $10^{7}
L_{\odot}$) in the ``halo'' term, we see that M33 lies even further
from the model lines.   How do we interpret this information?  A value
of $f_d$ = 1 implies that any satellite galaxy has been completely
destroyed and contributed all of its material to the halo.  The MW and
M31 data seem to favour a value of $f_d$ = 1, which is of course
inconsistent with observations (e.g., M31's latest tally is up to 29;
\citealt{2011ApJ...732...76R}, \citealt{2011ApJ...742L..14S},
\citealt{2011ApJ...742L..15B}).  If the M33 candidate halo fraction is closer to
the upper bounds, then it also appears that the models underestimate
the halo fraction for lower-luminosity galaxies. 

\section{Summary}
\label{summary}

We use Pan-Andromeda Archaeological Survey (PAndAS) data to identify
RGB candidate stars in the regions unrelated to the disk and extended
disk substructure surrounding the disk.  Contamination from 
both Milky Way foreground stars and misidentified background galaxies
is subtracted.  We reveal a new component centred on M33 that has a
low luminosity.  With such a weak signal, measurements are not
well constrained by our data.  However, it appears that this component
has an exponential scale length is of order $r_{exp} \sim $ 20 kpc, a
photometric metallicity of around [Fe/H] $\sim$ -2 dex, a
luminosity range of a few percent of M33's total host luminosity, and
is azimuthally asymmetric.  More observations and deeper photometry
are required to better determine the detailed structure of the stellar
populations. 

If this feature is truly a halo, it provides
support that stellar halos are a ubiquitous component of all galaxies,
built through the hierarchical merging predicted in $\Lambda$CDM
cosmology.


\acknowledgements
We thank the anonymous referee who provided useful comments that
helped to improve the paper.  RC and WEH thank the Natural Sciences
and Engineering Research Council 
of Canada for financial support.  GFL thanks the Australian Research
Council for support through his Future Fellowship (FT100100268) and
Discovery Project (DP110100678).  RI gratefully acknowledges support
from the Agence Nationale de la Recherche though the grant POMMME (ANR
09-BLAN-0228). Based on observations obtained with MegaPrime/MegaCam,
a joint project of CFHT and CEA/DAPNIA, at the Canada-France-Hawaii
Telescope (CFHT) which is operated by the National Research Council
(NRC) of Canada, 
the Institute National des Sciences de l'Univers of the Centre
National de la Recherche Scientifique of France, and the University of
Hawaii.  RC would like to thank the CFHT staff for much support. 
This research has made use of the NASA/IPAC Extragalactic Database
(NED) which is operated by the Jet Propulsion Laboratory, California
Institute of Technology, under contract with the National Aeronautics
and Space Administration. Thanks to Mike Rich for providing useful
comments that improved this paper.  


\bibliography{adsbibliography.bib}


\begin{deluxetable}{crrrrr}
  \tablecaption{The number of stellar objects located in each annulus
    shown in Figure \ref{frames}, and in each region shown in Figure
    \ref{cmds}. The sixth column includes all points out to a 
    radius of $r \lesssim$ 3.75 degrees. \label{numbers}}
  \tabletypesize{\scriptsize}
  \tablewidth{0pt}
  \tablehead
  {
\colhead{ }&\multicolumn{5}{c}{Annuli (degrees)}\\
\colhead{Region}&\colhead{0-1}&\colhead{1-2}&\colhead{2-3}&\colhead{3-3.75}&\colhead{All}
  }
  \startdata
M33 RGB&               83394&   7597&   7235&   3472&  101698\\
MW Disk&                9064&  15431&  26326&  13716&   64537\\
MW Halo&                9894&   4309&   6784&   3479&   24466\\
Total within annulus& 595878& 163178& 243949& 148885& 1151890\\
  \enddata
\end{deluxetable}

\begin{deluxetable}{cccccccc}
  \tablecaption{The fit parameters for the exponential model fits in Section
    \ref{radial_profile} shown in Figure
    \ref{rad_dens_background_uncorrected_corrected}.  Each column
    shows the quantity first in measured units, then in physical
    units in parentheses.  Units for each column are shown as
    footnotes.  The uncertainties on each parameter are shown under
    the columns labelled with $\Delta$. \label{fit_params}}
  \tabletypesize{\scriptsize}
  \tablewidth{0pt}
  \tablehead
  {
    \colhead{S/N}&\colhead{$\Sigma_{0}$ $^{a}$}&\colhead{$\Delta \Sigma_{0}$ $^{a}$}&\colhead{$r_{0}$ $^{b}$}&\colhead{$\Delta r_{0}$ $^{b}$}&\colhead{$\Sigma_{bg}$ $^{c}$}&\colhead{$\Delta \Sigma_{bg}$ $^{c}$}&\colhead{$\chi^{2}$}\\
  }
  \startdata
  15&233 (3.7)&189 (3.0)&1.0 (14)&0.6 (8)&365 (5.7)&14 (0.2)&1.2\\
  20&163 (2.6)&102 (1.6)&1.4 (20)&1.1 (16)&360 (5.6)&23 (0.4)&1.1\\
  25&158 (2.5)& 83 (1.3)&1.5 (21)&1.3 (18)&355 (5.6)&28 (0.4)&1.1\\
  30&182 (2.9)&124 (1.9)&1.3 (18)&1.0 (14)&361 (5.7)&23 (0.4)&1.2\\
  \hline
  \multicolumn{8}{l}{$^{a}$ Counts degree$^{-2}$ ($10^{-9}$ L$_{\odot}$ kpc$^{-2}$).}\\
  \multicolumn{8}{l}{$^{b}$ Degrees (kpc).}\\
  \multicolumn{8}{l}{$^{c}$ Counts degree$^{-2}$($10^{-9}$ L$_{\odot}$ kpc$^{-2}$).}\\
  \enddata
\end{deluxetable}

\end{document}